\documentclass[useAMS,usenatbib]{mn2e}

\title[Dust trapping in gravitationally unstable discs]{Dust trapping by spiral arms in gravitationally unstable protostellar discs}
\author[Giovanni Dipierro, Paola Pinilla, Giuseppe Lodato and Leonardo Testi]{
Giovanni~Dipierro$^1$\thanks{E-mail: giovanni.dipierro@unimi.it}, 
Paola Pinilla$^2$,
Giuseppe Lodato$^{1}$ and
Leonardo Testi$^{3,4,5}$ \\
$^1$ Dipartimento di Fisica, Universit\`a  degli Studi di Milano, Via Celoria 16, 20133 Milano, Italy\\
$^2$ Leiden Observatory, Leiden University, P.O. Box 9513, 2300 RA Leiden, The Netherlands\\
$^3$ European Southern Observatory, Karl Schwarzschild str. 2, D-85748 Garching bei M\"unchen, Germany\\
$^4$ INAF-Osservatorio Astrofisico di Arcetri, Largo E. Fermi 5, I-50125 Firenze, Italy\\
$^5$ Excellence Cluster Universe, Boltzmann str. 2, D-85748 Garching bei M\"unchen, Germany}

\usepackage[T1]{fontenc}  
\usepackage{graphicx}
\usepackage{xcolor}
\usepackage{hyperref}
\hypersetup{colorlinks,% 
citecolor=black,% 
filecolor=black,% 
linkcolor=black,% 
urlcolor=black}
\usepackage{indentfirst}
\usepackage[bf, small]{caption}
\usepackage{subfigure}
\usepackage{aecompl}
\usepackage{xspace}
\usepackage{amsmath}
\usepackage{epsfig}
\usepackage{verbatim}
\usepackage{setspace}
\usepackage{amsfonts}
\usepackage{color}
\usepackage{url}
\usepackage{booktabs}
\usepackage{fancyhdr}
\usepackage{textcomp}
\usepackage{natbib}
\usepackage{multirow}
\usepackage{textcomp}
\newcommand {\apgt} {\ {\raise-.5ex\hbox{$\buildrel>\over\sim$}}\ }
\newcommand {\aplt} {\ {\raise-.5ex\hbox{$\buildrel<\over\sim$}}\ }

\begin{document}
\date{}
\maketitle
\begin{abstract}
In this paper we discuss the influence of gravitational instabilities in massive protostellar discs on the dynamics of dust grains. 
Starting from a Smoothed Particle Hydrodynamics (SPH) simulation, we have computed the evolution of the dust in a quasi-static gas density structure typical of self-gravitating disc. For different grain size distributions we have investigated the capability of spiral arms to trap particles. We have run 3D radiative transfer simulations in order to construct maps of the expected emission at (sub-)millimetre  and near-infrared wavelengths. Finally, we have simulated realistic observations of our disc models at (sub-)millimetre  and near-infrared wavelengths as they may appear with the Atacama Large Millimetre/sub-millimetre Array (ALMA) and the High-Contrast Coronographic Imager for Adaptive Optics (HiCIAO) in order to investigate whether there are observational signatures of the spiral structure.
We find that the pressure inhomogeites induced by gravitational instabilities produce a non-negligible dynamical effect on centimetre sized particles leading to significant overdensities in spiral arms. We also find that the spiral structure is readily detectable by ALMA over a wide range of (sub-)millimetre wavelengths and by HiCIAO in near-infrared scattered light for non-face-on discs located in the Ophiucus star-forming region. 
In addition, we find clear spatial spectral index variations across the disc, revealing that the dust trapping produces a migration of large grains that can be potentially investigated through multi-wavelenghts observations in the (sub-)millimetric. 
Therefore, the spiral arms observed to date in protoplanetary disc might be interpreted as density waves induced by the development of gravitational instabilities.
\end{abstract}

\section{INTRODUCTION}
The physical and chemical evolution of protostellar discs plays a crucial role in the planet formation process. 
According to the most widely accepted scenario of planet formation,  the ``core accretion'' model, 
in the high-density environment of a circumstellar disc, micron sized dust grains are expected to grow via collisional agglomeration to kilometre-sized planetesimals (\citealt{Testi2014}). 
In recent years, the investigation on the dust evolution in protostellar discs has been the subject of several theoretical (e.g. \citealt{Brauer2008}, \citealt{Birnstiel2010}) and laboratory studies (e.g. \citealt{Blum2000}, \citealt{BlumWurm}) aimed at understanding dust dynamics and its growth. The general picture that emerges from these investigations is that the growth from submicron sized particles to larger objects is a complex process that contains many physical challenges (\citealt{Zsom}, \citealt{Okuzumi2009}, \citealt{Birnstiel2010}, \citealt{LaibeGonza2013a}). % closely linked to the dust dynamics and the grain growth processes.

Since the dust-to-gas mass ratio in protostellar discs is much lower than unity (typical value of  $\sim10^{-2}$), the dynamics of dust particles is heavily affected by the aerodynamic drag force that arises from the velocity difference between dust particles and the surrounding gas. For standard disc geometries, the pressure in the circumstellar disc tends to decrease outwards leading the gas to orbit at sub-Keplerian velocities. On the other hand, since the dust is not affected by the gas pressure, a freely orbiting dust particle feels only centrifugal forces and gravity, and therefore, orbits at Keplerian velocity to a first approximation. As a result, the drag force removes the dust angular momentum leading to an inward drift at a rate that depends on the size of particles \citep{Weidenschilling1977}. The radial velocity induced by gas drag is high enough to produce a rapid depletion of mm-sized particles in the outer disc before they can become large enough to decouple from the gas and, thus, preventing the growth of planetesimals required for the formation of the planetary cores. This is what is known as \emph{radial drift barrier}, and it still represent one of the main unsolved problem of the early phases of planet formation (\citealt{Weidenschilling1977}, \citealt{Nakagawa1986}, \citealt{Klahr}, \citealt{Brauer2008}). 

In addition, the growth of solid particles has to circumvent another hurdle closely linked to the sticking efficiency of colliding dust grains. It is expected that, while low-velocity collisions result in grain growth, high-velocity impacts between particles are destructive (\citealt{Poppe}, \citealt{Blum2000}, \citealt{Blum2006}). From laboratory and numerical studies it has been shown that, as particles become larger, they tend to collide at higher impact velocities.
Therefore, it is expected that the velocities of mm-sized objects in the outer disc are larger than the critical threshold for sticking, so that collisions lead to fragmentation and thus prevent dust particles from forming larger bodies (\citealt{Brauer2008}). As a result, the growth of dust particles toward meter sizes needs to overcome two obstacles: the rapid depletion of material due to the radial drift and the particle fragmentation produced by destructive collision. 

From an observational point of view, the evidence that grains in discs are significantly larger than those in the diffuse interstellar medium (ISM) has been obtained from a variety of observational techniques (see the reviews by \citealt{NattaTestiCalvetHenning2007} and \citealt{Testi2014}). %The level of growth evaluated by observations appears to be solid and consistent with theoretical expectations. 
Multi-wavelenght  (sub-) mm observations (e.g., \citealt{Testi2001,Testi2003}, \citealt{Ricci2010}, \citealt{Guilloteau}, \citealt{Perez}) of protostellar discs have shown that grain growth is efficient enough to quickly produce mm-sized particles which are retained in the outer disc for a relatively long time. Therefore, how to theoretically explain the retention of mm-sized particles in the outer regions of disc remains an open question. 

Several ideas of plausible mechanisms aimed at slowing down the rapid inward drift have been proposed, such as vortices, planet-disc interactions, snowlines, zonal flows or dead zones (e.g. \citealt{Klahr}, \citealt{Youdin2002}, \citealt{Johansen2009,Johansen2011}, \citealt{Pinilla}). 
The common feature of these mechanisms is to create long-lived radial and/or azimuthal inhomogeneites in the gas density structure in order to ``trap'' particles into the overpressure regions as discussed initially by \citet{2003ApJ...583..996H,2003ApJ...598.1301H}. %In detail, since the gas pressure gradient changes from positive to negative on one side of the inhomogeneity to negative on the other, the gas orbital velocity changes from super- to sub-Keplerian causing dust grains to drift toward the density/pressure maxima.
Some observational support for particle trap has been obtained from recent high angular resolution observations at mm-wavelength using ALMA that suggested the presence of a particle trap  by an anti-cyclonic vortex that creates dust segregation of small and large particles, as for example the case of IRS~48 \citep{2013Sci...340.1199V}.
%The macroscopic features in the dust distribution induced by the occurrence of these mechanisms can be observationally and unambiguously identified through high angular resolution observations. In this context, new and upgraded observing facilities, such as ALMA, are offering an unprecedented opportunity to put constraints on theoretical models.  In this context, \citet{2013Sci...340.1199V} have clearly revealed with ALMA the dust segregation of large particles in the outer part of a transitional disc due to an azimuthal inhomogeneity in the gas density induced by the presence of an embedded protoplanet.

In this paper we discuss the influence of the spiral density waves induced by gravitational instabilities (GI) on the dust dynamics. 
%It has been clearly recognized that the development of GI in protostellar discs plays a crucial role for their dynamical evolution. 
It is believed that in the early stage of star formation (Class 0/Class I objects), the disc could be massive enough to have a non-negligible dynamical effect on the evolution of the overall system. The disc self-gravity may affect the disc dynamics through the propagation of density waves that lead to the formation of a prominent spiral structure (\citealt{LodatoRice2004,LodatoRice2005}, \citealt{Durisen}, \citealt{CossinsLodatoClarke2009}, \citealt{Forgan2011}). In this context, it is expected that the density inhomogeneities induced by the development of GI affect both the dynamics and the growth of the dust grains. 
Recent work has already suggested that grain growth may be accelerated by concentrating the grains in the peak of the self-gravitating structures, possibly overcoming the issue of the rapid grain migration (\citealt{RiceLodato2004,RiceLodato2006}, \citealt{Gibbons1, Gibbons2}). Moreover, the dust retention in spiral arms implies low relative velocities between particles, which suggests that collisions are more likely to be constructive (\citealt{Gibbons1, Gibbons2}). 

In addition, the dust migration towards spiral arms can produce macroscopic features that can be identified through high angular resolution observations. 
Spiral and non-axisymmetric structures in protostellar discs have been observed recently in transitional discs by imaging the distribution of scattered light at near-infrared wavelengths (\citealt{GradyMuto2012}, \citealt{Garufi}, \citealt{GradyMuto2013}) and by high-resolution ALMA observations of molecular line emission (\citealt{Fukagawa}, \citealt{Christiaens2014}). For such evolved discs, one generally expects the disc not to be massive enough to be self-gravitating, and thus the more common explanation for the origin of the spiral is the dynamical interaction with an embedded planet. Nonetheless, it is worth remarking that disc mass estimates suffer from systematic errors, mostly due to uncertainties in the dust opacity and the assumption of a constant dust-to-gas ratio. Therefore, it is still debatable if these observed spirals are or not produced by planet-disc interactions \citep{Juhasz}. 
Moreover, \citet{Dipierro2014} (see also \citealt{CossinsLodatoTesti2010}) have found that, assuming that the dust is perfectly mixed with the gas, spiral structures of a variety of non-face-on self-gravitating circumstellar discs models with different properties in mass and radial extension are readily detectable by ALMA over a wide range of wavelengths. 
One of the aims of the present work is to improve the observational predictions of self-gravitating circumstellar discs presented in \citet{CossinsLodatoTesti2010} and \citet{Dipierro2014} by including the treatment of dust dynamics (in particular, by relaxing the hypothesis that gas and dust are perfectly mixed) and a more detailed procedure for the generation of emission maps.

In this work, by combining the results of hydrodynamic dust and gas simulations with 3D Monte Carlo radiative transfer calculations (using RADMC-3D), we study if the occurrence of local pressure maxima induced by GI can trap dust particles and investigate the detectability of these inhomogeneites at near-infrared and (sub-)millimetre  wavelengths.

The paper is organized as follows: in Section \ref{sec:models} we describe the details of the hydrodynamic simulations. In Section \ref{sec:results} we describe the results of the simulations of the dust dynamics and present the observational predictions of our disc models. Finally, in Section \ref{sec:discuss} we discuss the significance of our results.

\section{GAS AND DUST DISC MODELS}
\label{sec:models}
The simulations presented here calculate the evolution of the dust surface density in a quasi-static gaseous self-gravitating disc. 
The system comprises two components discs: a gas disc, characterized by a spiral surface density structure induced by GI, and a dusty disc that evolves under the action of drag forces and turbulent mixing induced by the gas-dust aerodynamical coupling. The self-gravitating gas disc model has been taken from a snapshot of a three-dimensional Smoothed Particle Hydrodynamics (SPH) simulation performed by \cite{LodatoRice2004}. This model can be considered as representative of a self-gravitating gas disc in a quasi-steady state characterized by a spiral surface density structure produced by a combined effect of self-gravity and cooling.
We compute the dust surface density evolution in such fixed gaseous background in the radial direction along different azimuthal cuts across the disc. Afterwards, starting from the resulting surface density structure we create 3D models using a representative estimator for the tickness of the dusty disc. We carry out 3D radiative transfer simulations in order to create the expected emission maps at (sub-)millimetre  and near-infrared wavelengths of our models which are used to simulate realistic observations.

\subsection{Gas disc model}
\begin{figure}
%\begin{minipage}{\textwidth}
\centering
\includegraphics[scale=0.48]{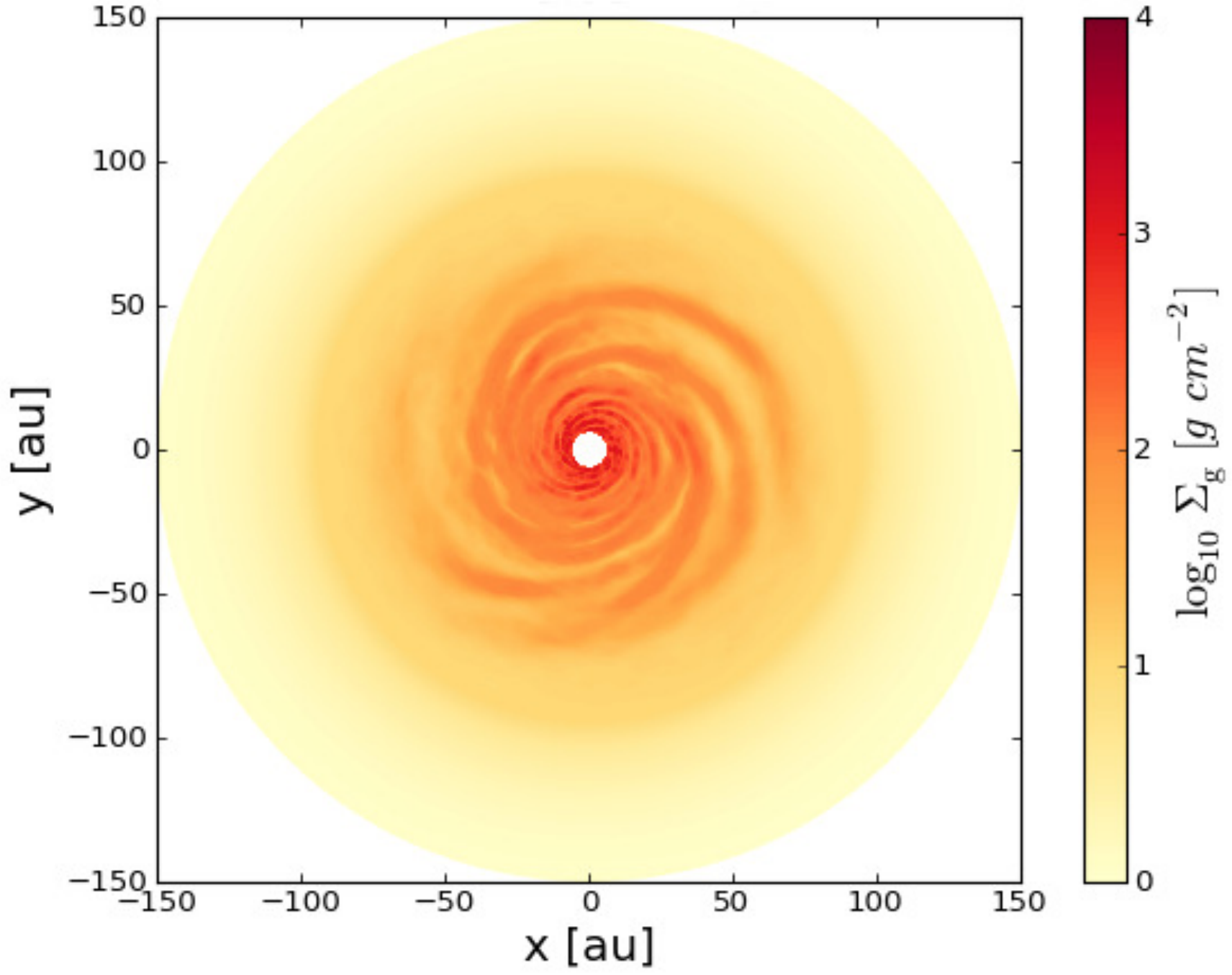}
\caption{Gas surface density structure of a $0.25 M_{\odot}$ disc around a $1M_{\odot}$
star that has reached a long-lived, self-gravitating state.}
\label{img:plot_density}
%\end{minipage}
\end{figure}
The gas disc model used as an input for the dust simulations has been taken from \cite{LodatoRice2004}. They have performed three-dimensional SPH simulations of self-gravitating discs evolving under an analytically prescribed cooling function. The system comprises a central star modelled as a point mass $M_{\star}$  surrounded by a gaseous disc of mass $M_{\rm{disc}}$ which extends from $R_{\rm{in}}=0.25$ to $R_{\rm{out}}=25$ in code units.
%The disc extends from $R_{\rm{in}}=0.25$ to $R_{\rm{out}} = 25$ code units and, in these units, one dynamical time-scale at $R=1$ is equal to $2 \pi$ code units. Therefore, one orbital period at the outer edge of the disc is roughly equal to $800$ time units. 
The simulation used in this work was performed using a value for the disc-to-central object mass ratio $q=M_{\rm{disc}}/M_{\star}=0.25$. The disc is characterized initially by a surface density profile $\Sigma \propto R^{-1}$ while the initial temperature profile is $T \propto R^{-1/2}$. %With these initial surface density and temperature profiles, the stability parameter $Q$ (see \cite{LodatoRice2004}) decreases with increasing radius: $Q \propto c_s \Omega/\Sigma \propto R^{-3/4}$. 
%The temperature is normalized such the disc is initially gravitationally stable and, therefore, the sources of gravitational heating is temporary turned off.
%The initial distribution of particles is such that the vertical density profile is gaussian with a scale height $H=c_s/\Omega$ where $c_s$ is the sound speed and $\Omega$ is the angular velocity. It is worth noting that the vertical density profile for a self gravitating disc in dynamical equilibrium is not rigorously gaussian (see \cite{BertinLodato1999}). However, any variation from dynamical equilibrium is washed out on the dynamical timescale $\sim \Omega^{-1}$.
Regarding the thermal aspect of disc evolution, the heating is governed by both $P\rm{d}V$  work and viscous dissipation while the cooling is implemented using a simple cooling law given by \citep{2001ApJ...553..174G}:
\begin{equation}
\frac{d u_{\rm{i}}}{dt}=-\frac{u_{\rm{i}}}{t_{\rm{cool},\rm{i}}}\, ,
\label{tcool}
\end{equation}
where $u_{\rm{i}}$ is the specific internal energy and $t_{\rm{cool},i}$ is the cooling time associated with the $i^{th}$ particle. The latter is determined using the simple parametrization $\Omega_{\rm{i}} \, t_{\rm{cool},\rm{i}}=\beta_{\rm{cool}}$, where $\beta_{\rm{cool}}$ is held constant and equal to $7.5$.  %This form of cooling has been used extensively in simulations of discs in various context, especially in studying the properties of the density perturbation due to the onset of gravitational instability and the rate at which the disc cools (e.g., \cite{CossinsLodatoClarke2009}). 
\begin{figure}
%\begin{minipage}{\textwidth}
\centering
\includegraphics[scale=0.475]{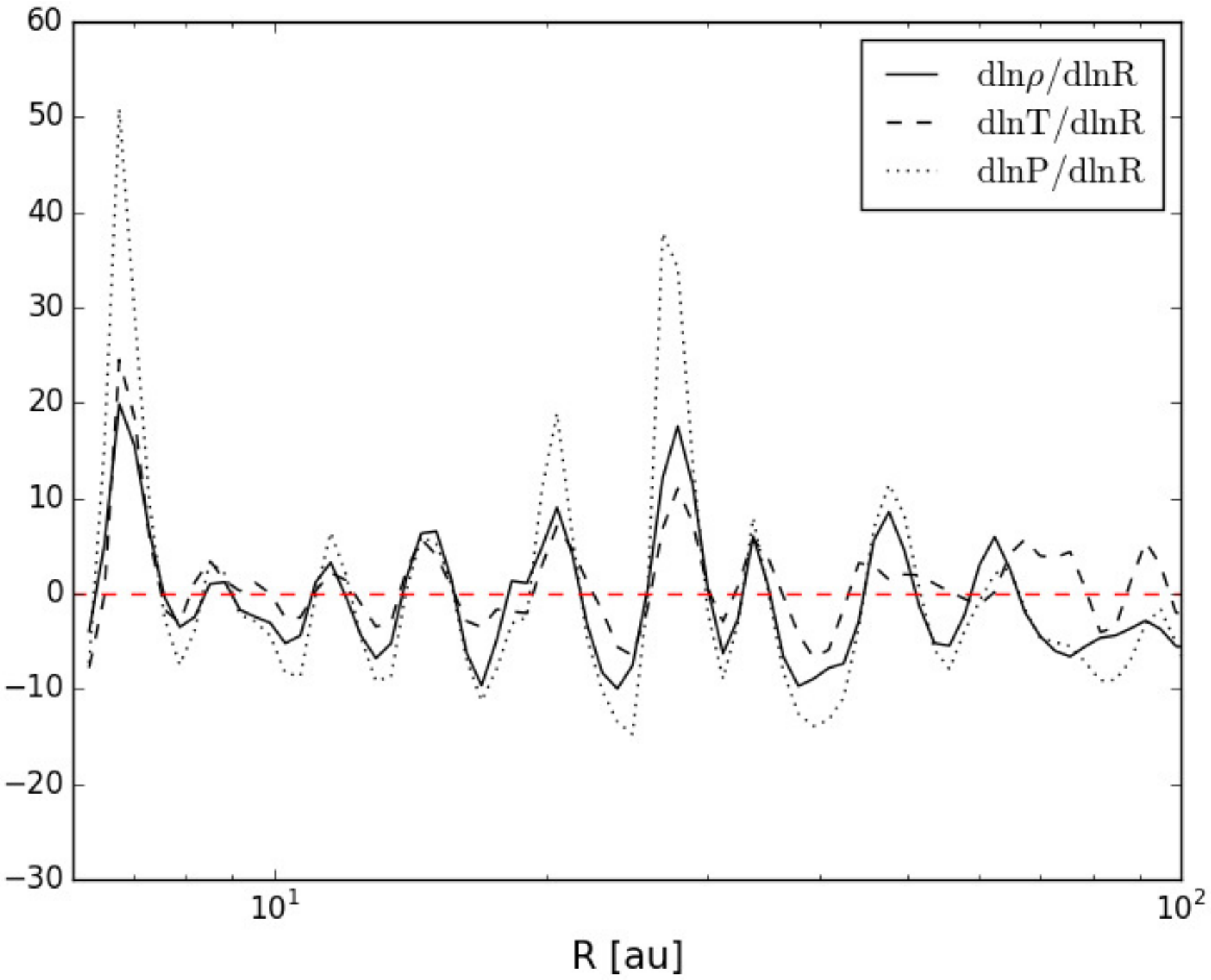}
\caption{Gas volume density, temperature and pressure gradient along a slice across the disc.}
\label{img:gradient}
%\end{minipage}
\end{figure}
All simulations described above are essentially scale-free. In order to consider a gas disc model with realistic properties in mass and radial extension, we modify the length and the mass scale of the data output of the SPH simulations such that the radial extent of the disc is $150$ au with a disc mass equal to $0.25\, \rm{M_{\odot}}$. 
As shown in \cite{LodatoRice2004}, the initial phase of disc evolution is governed by a rapid cooling until the gravitational instability becomes effective, providing a source of effective heating through both compression and shocks. In approximately one thermal time, the disc settles into a quasi-steady state developing spiral density waves that propagate across the disc.
Once the thermal equilibrium state is reached, the gas disc is characterized by a spiral structure (shown in Fig. \ref{img:plot_density}) where the heating generated through GI balances the external cooling. This structure remains quasi-steady, evolving secularly on the viscous timescale ($10^5$-$10^6$ yrs in this case). Individual spiral arms are recurrent structures evolving on the dynamical timescale ($\sim 10^3$ yrs in the outer disc).

Fig. \ref{img:gradient} shows the volume density, temperature and pressure gradient along a slice across the disc.  It can be noticed that all the gradients are characterized by a similar trend and are null at the same radial positions.
Therefore, the overdensity regions, which correspond to the arm regions, are characterized by relatively higher temperatures while in the interarm region the temperature is lower (see also Fig. 3 of \citealt{Dipierro2014}). This is due to the increase in the pressure caused by a shock driving a compression of gas and, as a consequence, a growth of the temperature. In the interarm regions, the increase in the pressure provides an expansion of gas parallel to the shock and, as a consequence, a decrease of the temperature \citep{CossinsLodatoClarke2009}. Moreover, since the pressure depends on the volume density and the sound speed, the peaks of the pressure gradient are higher than those related to the temperature and volume density profile.

\subsection{Dust evolution model}
\label{sec:evol_model} 
To describe the evolution of the dust due to drag forces and turbulent diffusion it is convenient to define how well coupled the particles are to the gas. The gas-dust aerodynamical coupling is expressed by the Stokes number, usually denoted by St, defined as the ratio of the stopping time of a particle and the local dynamical timescale ($\Omega^{-1}$). 
Generally, there are four different regimes for the dust-to-gas aerodynamical coupling depending on the value of the Reynolds number and the ratio between mean free path $\lambda_{\rm{mfp}}$ of the gas molecules and the dust particle size, $a$. In the Epstein regime, for $\lambda_{\rm{mfp}}/a\geq4/9$, the Stokes number for particles close to the disc mid-plane, is given by \citep{Birnstiel2010}:
\begin{equation}
\mathrm{St}=\frac{a \rho_s}{\Sigma_{\rm{g}}}\frac{\pi}{2} \, ,
\label{eq:stokes}
\end{equation}
where $\Sigma_{\rm{g}}$ is the gas surface density and $\rho_s$ is the internal density of dust grains.
The radial velocity of dust particles can be determined by self-consistently solving the radial and azimuthal components of the momentum equation \citep{Weidenschilling1977}:
\begin{equation}
u_{\rm{r}}=u_{\rm{drag}}+u_{\rm{drift}}=\frac{u_{\rm{g,r}}}{1+\rm{St}^2}+\frac{1}{\rm{St}+\rm{St}^{-1}}\, \frac{1}{\rho_{\rm{g}} \Omega} \frac{\partial P}{\partial R} \, ,
\label{eq:rad_vel}
\end{equation}
where $\Omega$ is the Keplerian angular velocity, $P$ is the pressure and $\rho_{\rm{g}}$ is the gas volume density.
The first term is the gas drag term which is produced by the coupling of the dust particle to gas moving with a radial velocity of $u_{\rm{g,r}}$ given by viscous evolution.
The second term is the radial drift caused by the angular momentum exchange between dust particles and the non-Keplerian orbital motion of the gas. If the Stokes number is much smaller than unity, the dust particle is strongly coupled to the gas, which implies that the gas and dust share the same motion. On the other hand, for $\rm{St}>>1$, the particles are unaffected by the gas and consequently move on Keplerian orbits to a first approximation. Particles in the intermediate size range, i.e. $\rm{St}\sim1$, have large drift velocities and are expected to experience the highest concentration in pressure maxima.

Since the gas is expected to be turbulent, the dust is mixed by the gas due to aerodynamical coupling. The turbulent transport depends on the gas diffusion $D_{\rm{g}}$, which is usually described by the canonical turbulence prescription \citep{Shakurai}.
%In this prescription the kinematical viscosity is expressed following the assumptions that the magnitude of viscosity driven by local turbulence can be roughly given by $\nu \sim v l$, where $l$ is the typical size of the largest eddies ($l\aplt H$) and $v$ is the typical turbulent velocity which is
%expected to be lower than the local sound speed. 
The dust diffusivity induced by the gas-dust aerodynamical coupling is defined as \citep{Youdin}: 
\begin{equation}
D_{\rm{d}}=\frac{D_{\rm{g}}}{1+\rm{St}^2}=\frac{\alpha \langle c_{s}\rangle H_{\rm{g}} }{1+\rm{St}^2}  \, ,
\label{eq:diff}
\end{equation}
where $H_{\rm{g}}$ is the gas scale height, $\langle c_s \rangle$ is the azimuthal average of the sound speed and $\alpha$ is an unknown dimensionless parameter which embodies all the uncertainties
about the nature of the viscous mechanism. Since the diffusion time scale is longer than the lifetime of individual spiral features in self-gravitating discs ($\sim \Omega^{-1}$), the dust diffusivity is computed using the azimuthal average of the sound speed.

Generally, it is expected that a gravitationally unstable disc cannot be treated as a viscous disc because gravity can affect the transport mechanism over long distances.
In this context, a number of numerical simulations have been recently carried out with the aim to test the locality of transport provided by GI. \citet{LodatoRice2004}, \citet{CossinsLodatoClarke2009} and \citet{Forgan2011}, using an $\alpha$-like approach, have compared the dissipation power provided by gravitational instability with the expectations based on a viscous theory of disc.
They have found that for disc to star mass ratio $M_{\rm{disc}}/M_{\star} \aplt 0.25$, the gravitational instability is well described using a viscous approach. 

Encompassing all the systematic and random motions described above, the time evolution for the dust surface density $\Sigma_{\rm{d}}^{\rm{i}}$ of the $\rm{i^{th}}$ dust species with size $a^{\rm{i}}$ can be described by an advection-diffusion equation (\citealt{Brauer2008}, \citealt{Birnstiel2010}):
\begin{equation}
\frac{\partial \Sigma_{\rm{d}}^{\rm{i}}}{\partial t}+\frac{1}{R}\frac{\partial}{\partial R} \left [ R\, \biggl(\Sigma_{\rm{d}}^{\rm{i}} \, u_{\rm{r}}^{\rm{i}} - D_{\rm{d}}^{\rm{i}} \, \Sigma_{\rm{g}}\, \frac{\partial}{\partial R} \biggl (\frac{\Sigma_{\rm{d}}^{\rm{i}}}{\Sigma_{\rm{g}}} \biggr )\biggr ) \right ]=0 \, ,
\label{eq:fulleq}
\end{equation}
where, as previously described, the diffusion coefficient (see Eq. \ref{eq:diff}) and the radial velocity (see Eq. \ref{eq:rad_vel}) depends on the Stokes number and therefore on the particle size.

In this work, the dust evolution for each grain size in our sample is computed using the numerical method adopted in \citet{Birnstiel2010}. All the simulations performed here are evolved for one outer orbital period: $10^3$ yrs. Such timescale is long enough to ensure that the drag forces and turbulent diffusion affect the dust density structure but short enough that we do not expect a significant evolution of the spiral structure. 
We cut the disc into 720 radial directions. For each cut, we compute the gas pressure needed in Eq. \ref{eq:rad_vel}. We then solve, for each cut, Eq. \ref{eq:fulleq} using 100 radial cells, so as to obtain the 2D surface density of the dust.
%We model the disc using 100 and 720 radial and azimuthal cells, respectively.  Thus, by performing 720 simulations of the dust density evolution in the radial dimension along different azimuthal directions, we combine the results in order to obtain the overall dust evolution.

\subsection{Initial conditions}
The evolution of the dust density and the resulting observational signatures depend on the initial conditions of the simulations.  Generally, the distribution of grain size in a protostellar disc is affected by the grain growth and fragmentation processes which in turn are closely linked to the relative velocities between colliding particles. In this work, we assume that the dusty disc has settled into a coagulation-fragmentation steady state. Since the self-gravitating phase of protostellar disc is more likely to occur in the early stage of star formation, such initial condition for the dust density ditribution might appear simplistic. However, recent observational results have shown that large dust aggregates can form during the disc formation stage in the infalling envelope \citep{Miotello}.
The distribution of grain size is assumed to be a standard power-law size distribution:
\begin{equation}
n(a)\propto a^{-q} \qquad \mathrm{for} \quad a_{\rm{min}}<a<a_{\rm{max}} \, ,
\label{grainsizedistrib}
\end{equation}
where $a_{\rm{min}}$ and $a_{\rm{max}}$ represent the minimum and maximum size of the grains.  The dust size distribution is modelled using 100 sizes and it is normalized such that the total initial dust density is one per cent of the gas density in each cell of the grid.  The minimum size of the grains is assumed to be equal to  $0.1\, \rm{\mu m}$.

Regarding the value of $a_{\rm{max}}$, \citet{Birnstiel2010} have found that an acceptable estimator for the maximum grain size of the dust distributions can be computed by assuming the random turbulence motion as the only source of relative motions (for a detailed discussion of the turbulent relative velocities see \citealt{OrmelCuzzi2007}). In this context, \cite{RiceLodatoArmitage2005} have found that the turbulent diffusion induced by GI for the gas disc model used here, is characterized by a ``gravoturbolent'' parameter $\alpha=0.05$. Since the turbulence parameter related to GI is reasonably larger than the typical turbulence parameters, the previously mentioned assumption for the maximum grain size of the dust size distribution can be considered a valid approximation.
The value of the maximum grain size is therefore computed by equating the relative velocities between colliding particles induced by ``gravoturbolent'' diffusion and the threshold velocity for fragmentation (\citealt{Birnstiel2010}). In the Epstein regime, this is given by:
\begin{equation}
a_{\rm{max}}= \frac{4 \langle \Sigma_{\rm{g}} \rangle}{3\pi \alpha \rho_s}\frac{v^2_{\rm{frag}}}{\langle c_s \rangle^2}\, ,
\label{eq:amax}
\end{equation}
where $v_{\rm{frag}}$ is the fragmentation threshold velocity and $\langle \Sigma_{\rm{g}} \rangle$ is the azimuthal average of the gas surface density. Similarly to the computation of the dust diffusivity (Eq. \ref{eq:diff}), since the typical grain growth timescales is longer than the lifetime of individual spiral features in self-gravitating discs (\citealt{Brauer2008}, \citealt{Okuzumi2011}), $a_{\rm{max}}$ is computed using the azimuthal average of the gas surface density and sound speed.
Regarding the value of the fragmentation velocity, it has been shown through theoretical and experimental studies that it depends on dust grains material properties. Typically, the fragmentation velocity for silicates dust aggregates is of the order of a few $\rm{m~s}^{-1}$ \citep{BlumWurm} while icy aggregates are able to grow at collisions with velocities up to 50 $\rm{m~s}^{-1}$ (\citealt{Wada}, \citealt{2015ApJ...798...34G}).

Moreover, the dust size distribution is characterized by the value of the power-law exponent q that depends on the dust dynamics and the processes of grain coagulation and fragmentation that occur in the disc. While the typical value in the ISM is $q_{\rm{ism}} = 3.5$ \citep{Mathis1977}, in a protoplanetary disc if coagulation (fragmentation) processes dominate over fragmentation (coagulation),  a lower (higher) value for $q$  than the ISM one is expected. 

In order to cover a wide range for the initial dust size distributions, we perform simulations using different values for the fragmentation threshold velocity and the power-law exponent. We consider two values for the fragmentation threshold velocity: $v_{\rm{frag}}=[10, 30] \,\rm{m~s}^{-1}$ and three values for the power-law exponent: $q=[3.0, 3.5, 4.0]$.

\subsection{Monte Carlo radiative transfer simulation}
\label{subsec:montecarlo}
We calculate the expected emission maps of the self-gravitating disc model adopted here using the 3D radiative transfer code RADMC-3D\footnote{http://www.ita.uni-heidelberg.de/~dullemond/software/radmc-3d}. We focus on images at H-band (1.65 $\rm{\mu}$m) and ALMA band 6 (220 GHz), band 7 (345 GHz), band 8 (460 GHz) and band 9 (680 GHz). Our aim is to compute HiCIAO scattered light images in H-band polarized intensity and ALMA (sub-)millimetre  images in order to study the detectability of the peculiar spiral structure induced by GI both in scattering and thermal dust emission.

%We compute realistic HiCIAO and ALMA observations in order to trace opposite ends of the dust particle size distribution. In detail, since 
It is known that the scattering intensity is closely linked to the behaviour of the scattering phase function of dust grains. While small grains scatter photons uniformly, the scattering phase function of large grains is dominated by forward scattering. Therefore, since large grains scatter stellar photons mostly into the disc, tracing larger grains by analyzing scattering images of non-edge-on discs results challenging  \citep{Mulders}. Thus, polarized scattering images at short wavelengths trace small (0.1-10 $\rm{\mu}$m) dust grains at the surface of the disc (where stellar photons get absorbed or scattered).
At (sub-)millimetre  wavelengths, since the disc is generally optically thin in the vertical direction, the emission probes the mid-plane disc surface density in large grains (0.1-10 mm), due to their high opacity at these wavelengths (\citealt{2007prpl.conf..555D}, \citealt{2011ARA&A..49...67W}). 

In order to perform RADMC-3D simulation, we need to create 3D structures starting from the 2D dust density distributions obtained with the procedure described in Sect. \ref{sec:models}. In a protostellar disc the vertical structure of the dust is determined by the balance between the dust settling toward the disc mid-plane due to the gravity and the vertical turbulent stirring induced by the gas drag. 
The dust density profile of the $\rm{i^{th}}$ dust species with size $a^{\rm{i}}$ is assumed to be Gaussian:
\begin{equation}
\rho_{\rm{d}}^{\rm{i}}(z)=\frac{\Sigma_{\rm{d}}^{\rm{i}}}{h_{\rm{d}}^{\rm{i}}\sqrt{2 \pi}} \, exp \biggl({-\frac{z^2}{2 {h_{\rm{d}}^{\rm{i}}}^2}}\biggr) \, ,
\end{equation}
where the scale height $h_{\rm{d}}^i$ is computed assuming that the gravitational settling balances the turbulent diffusion (\citealt{Brauer2008}, \citealt{Birnstiel2010}):
\begin{equation}
h_{\rm{d}}^{\rm{i}}=H_{\rm{g}}(R) \, \rm{min} \Biggl(1,\sqrt{\frac{\alpha}{\rm{min}({St}^{\rm{i}},1/2)(1+{{St}^{\rm{i}}}^2)}}\Biggr) \, ,
\end{equation}
where, bearing in mind that the disc is in margina self-gravitating state, the gas scale height is given by \citep{CossinsLodatoClarke2009}:
\begin{equation}
H_{\rm{g}}(R)=\frac{\pi \Sigma_{\rm{g}}(R) R^3}{M_{\star}}\, .
\end{equation}
Having constructed the 3D distribution for all the dust species in our sample, we compute the temperature structure of the dust via thermal Monte Carlo simulations. The source of radiation is assumed to be the central star, located at the centre of the coordinate system, with $M_{\star}=M_{\odot}$, $R_{\star}=R_{\odot}$ and $T_{\rm{eff}}=6000$ K. The dust in our model consists of spherical composite porous grains with chemical abundances adopted in \citet{Ricci2010}. The absorption and scattering cross sections for each dust species in our sample are calculated using the routine developed by \citet{BohrenHuffman}. In addition, in order to compute near infrared scattered light images in the H-band, we include the full treatment of polarization by computing the scattering matrix for each dust species. 
% take into account the anisotropy of the scattering by computing the average cosine scattering angle that is used by RADMC-3D to calculate the scattering phase function in the Henyey-Greenstein approximation \citep{HenyeyGreenstein}.

Starting from the dust temperature structure of the model, the expected emission maps are computed via raytracing assuming that the disc is non-face-on (inclination angle of 30\textdegree).
The polarized intensity image (PI=$\mathrm{\sqrt{Q^2+U^2}}$) is calculated starting from the full Stokes vector (I,Q,U,V) computed by RADMC-3D.
The full-resolution (sub-)mm images directly produced by RADMC-3D are used as input sky models to simulate realistic ALMA observations using the Common Astronomy Software Application (CASA) ALMA simulator. The HiCIAO images are computed by convolving the full-resolution images in H-band with a measured HiCIAO point spread function taken from the ACORNS-ADI SEEDS Data Reduction Pipeline software \citep{Brandt}.
%The creation and analysis of the simulated ALMA images is carried out using the same approach adopted in \citet{Dipierro2014}. 
%All the source models are assumed to be non-face-on (inclination angle of 30\textdegree) and located in Ophiucus star-forming region (distance of 140 pc). 

\section{RESULTS}
\label{sec:results}
\begin{figure*}
\begin{minipage}{\textwidth}
\centering
\includegraphics[scale=0.48]{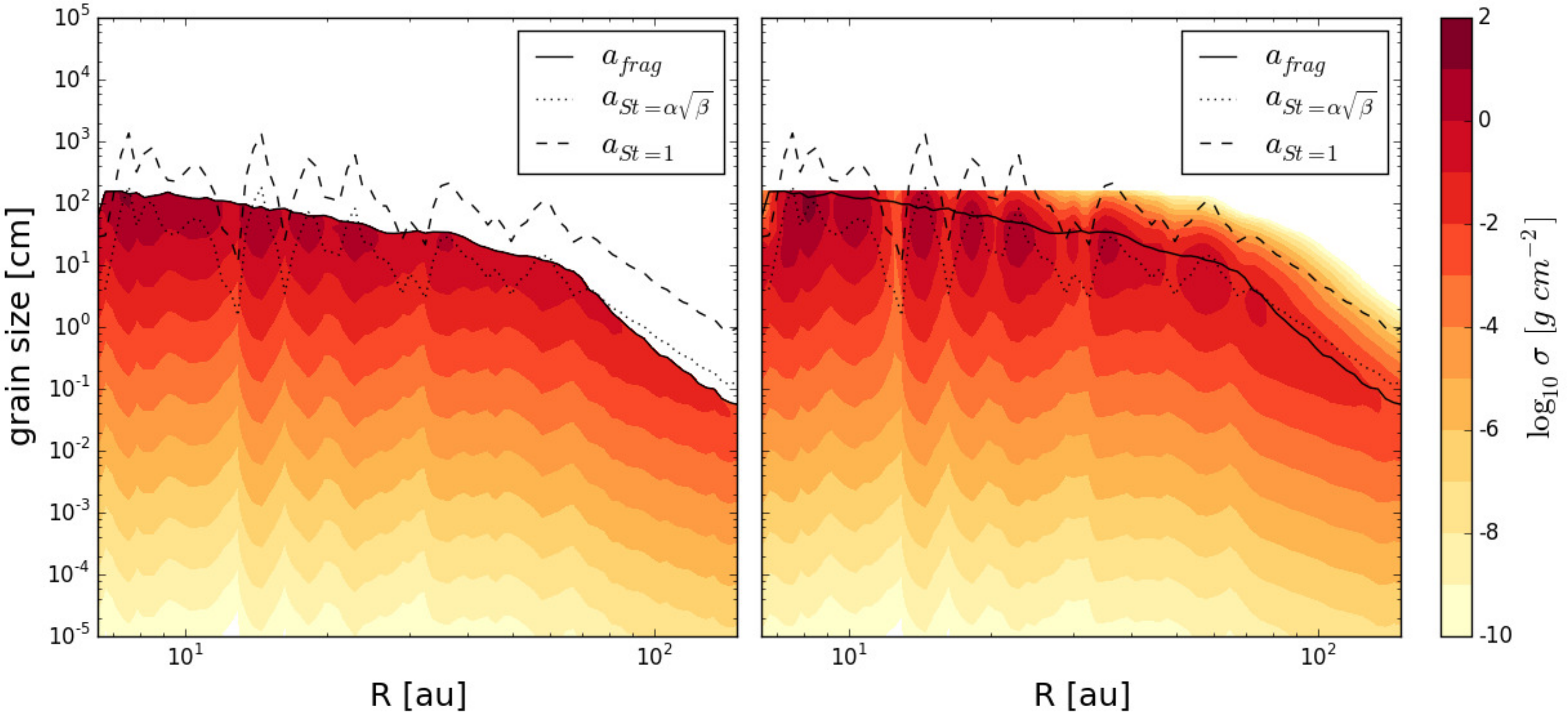}
\caption{Dust density distribution at the beginning (left) and at the end (right) of simulation ($t=10^3 \,\rm{yrs}$) along a slice across the disc. The simulation starts from an initial size distribution with exponent  $q=3.5$ and a fragmentation velocity equal to 30 $\rm{m~s}^{-1}$. The dashed line represents the particle size corresponding to a Stokes number of unity while the dotted line show the particle size where $St=\alpha \sqrt{\beta_{\rm{cool}}}$. The solid line shows the fragmentation barrier, i.e. the maximum value of the initial grain size distribution (see Eq. \ref{eq:amax}).}
\label{img:plot_cut}
\end{minipage}
\end{figure*}
In this section, we present results of the dust simulations and the observational predictions of the resulting disc models. We will focus on the level of the arm-interarm contrast induced by gas-dust coupling for small and large grains and how the dust evolution and observational signatures change with different initial grain size distributions.

\subsection{Dust density structure}
%In the following we present the results of numerical simulations of dust evolution in a non-axisymmetric gas density structure under the action of systematic and random motions induced by the gas-dust aerodynamical coupling. 
First, we show a representative set of results with a given set of model parameters. Figure \ref{img:plot_cut} shows the initial (left) and the final (right) dust density distribution along a slice across the disc from a simulation starting from an initial size distribution with exponent  $q=3.5$ and a fragmentation velocity equal to 30 $\rm{m~s}^{-1}$. The colour contours denote the dust surface density distribution $\sigma(a,R)$ per logarithmic size bin, defined through:
\begin{equation}
\Sigma_{\rm{d}} (R)=\int_{a_{\rm{min}}}^{a_{\rm{max}}(R)} \sigma(a,R) \, \mathrm{d}\, \mathrm{ln} a \, ,
\end{equation}
where $\Sigma_{\rm{d}}$ is the total dust surface density. The solid line shows the fragmentation barrier, i.e. the maximum value of the grain size distribution (see Eq. \ref{eq:amax}). The dashed line represents the particle size corresponding to a Stokes number of unity, which represent the size of particles that experience the fastest radial drift. Since the dust is in the Epstein regime, this line reflects the shape of the surface density, as expressed in Eq. \ref{eq:stokes} so that peaks in this quantity refer to the high density arms, while valleys indicate the interarm regions. The dotted line in Fig. \ref{img:plot_cut} shows the particle size corresponding to a Stokes number of $\alpha \sqrt{\beta_{\rm{cool}}}$, which, as described below,  represent the size above which the radial migration towards pressure maximum dominates over turbulent diffusion. 
It can be noticed that particles with Stokes number $\rm{St}\sim 1$ tend
to be affected by spiral density waves most effectively and exhibit the largest concentration in wave crests. The dust migration toward pressure maxima is negligible for smaller particles. This is due to the higher coupling between the dust and the slowly evolving gas and also because the diffusion is strong enough such that variations in the dust-to-gas ratio are being smeared out. 
In addition, since the maximum value of the initial grain size distribution decreases with radius, there is a strong density gradient of the largest particles between adjacent cells along the radial grid. Therefore, the gravoturbulent diffusion produces a fast migration of larger particles towards outer regions leading to an additional density enhancement in the spiral arm as can be easily noticed in the right panel of Fig. \ref{img:plot_cut}, where a large population of particles above the fragmentation limits is present. 
The motion towards outer regions induced by diffusion is more evident when the slope of maximum grain size increases (for $R>70$ au) because a higher number of particles diffuse towards the outer radius. However, it is expected that largest particles beyond the fragmentation threshold tend to experience disruptive collisions on longer timescales. 
\begin{figure*}
\begin{minipage}{\textwidth}
\centering
\includegraphics[scale=0.45]{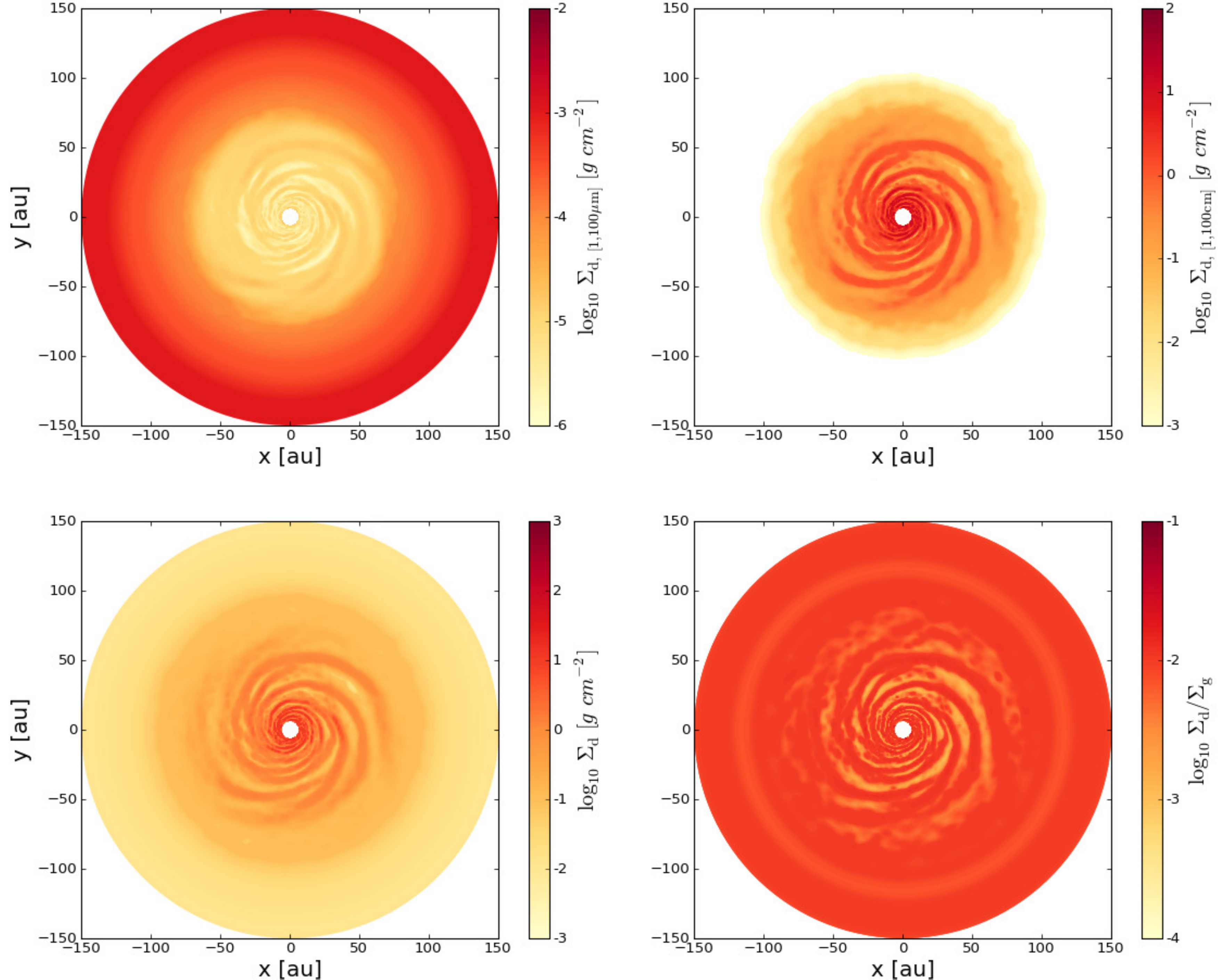}
\caption{The upper panels show the surface density at the end of simulation ($t=10^3 \,\rm{yrs}$) integrated in size in the range $[1\mu m,100\mu m]$ (left) and $[1 cm, 100 cm]$ (right). The lower panels show the total dust density distribution (left) and the dust-to-gas ratio (right) at the end of simulation.}
\label{img:plot_total}
\end{minipage}
\end{figure*}
\begin{figure*}
\begin{minipage}{\textwidth}
\centering
\includegraphics[scale=0.48]{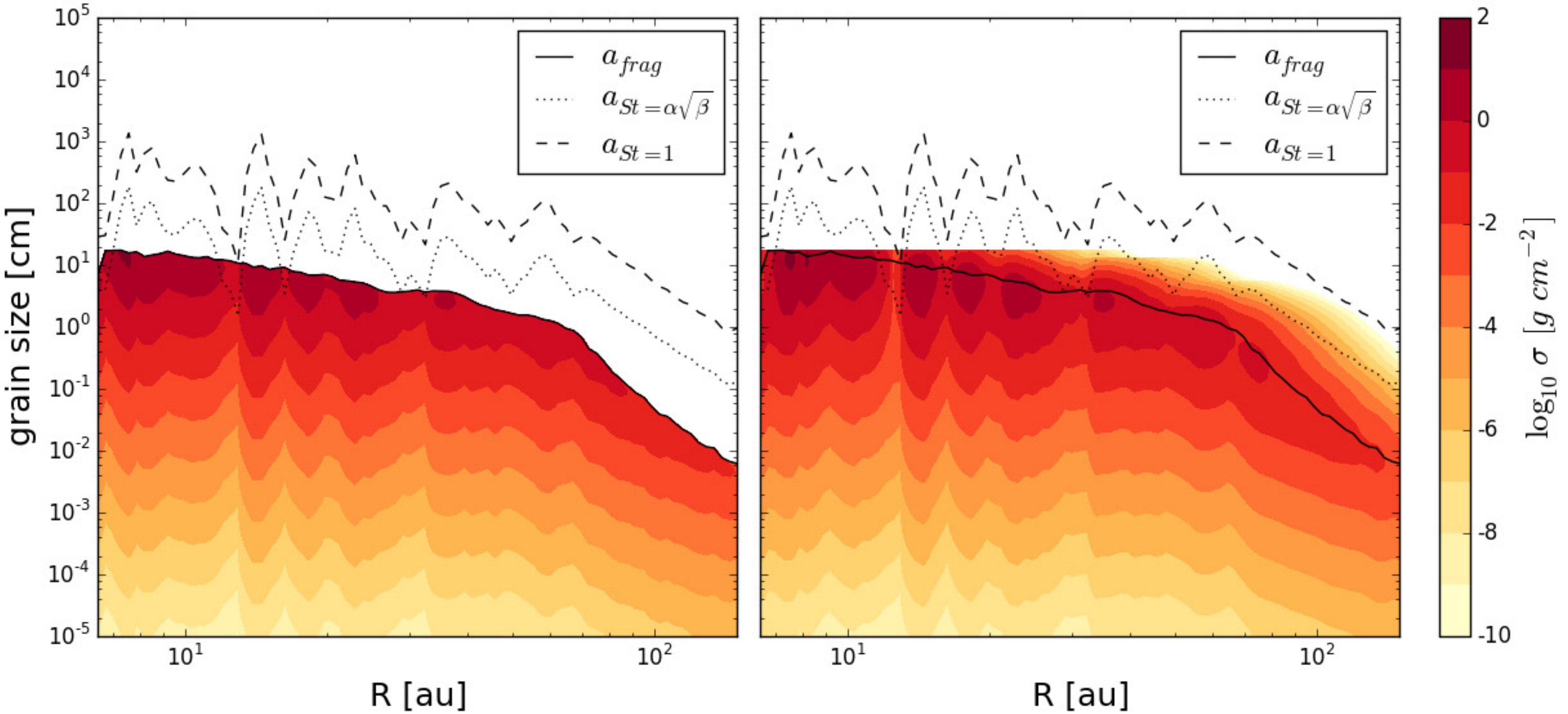}
\caption{Dust density distribution at the beginning (left) and at the end (right) of simulation ($t=10^3 \,\rm{yrs}$) along a slice across the disc. The simulation starts from an initial size distribution with exponent  $q=3.5$ and a fragmentation velocity equal to 10 $\rm{m~s}^{-1}$. The dashed line represents the particle size corresponding to a Stokes number of unity while the dotted line show the particle size where $St=\alpha \sqrt{\beta_{\rm{cool}}}$. The solid line shows the fragmentation barrier, i.e. the maximum value of the initial grain size distribution (see Eq. \ref{eq:amax}).}
\label{img:plot_cut_v1}
\end{minipage}
\end{figure*}

It is known that the dust concentration in pressure bumps is the result of the competition between the radial drift, that forces the particle to migrate towards pressure maxima, and the turbulent diffusion that tends to recover the dust density distribution into the initial condition where the dust-to-gas ratio is equal in the overall disc. In order to figure out the relative importance of advection and diffusion, we can compare the advection time scale, that can be estimated by the minimum time required to concentrate solid material in a pressure maximum: $t_{\rm{adv}}\sim \lambda/u_{\rm{drift}}$, and the diffusion time scale $t_{\rm{diff}}\sim \lambda^2/D_{\rm{d}}$, where $\lambda$ is the scalelength of the density inhomogeneities that, in a gravitationally unstable disc, is of the order of the disc scale height $H_{\rm{g}}$. The ratio of these timescale is given by:
\begin{equation}
\frac{t_{\rm{diff}}}{t_{\rm{adv}}}=\frac{\rm{St}}{\alpha}\frac{\Delta \Sigma_{\rm{g}}}{\Sigma_{\rm{g}}} \, ,
\end{equation}
where, since we have a surface density enhancement $\sim \Delta \Sigma_{\rm{g}}$ over a radial length scale  $H_{\rm{g}}$, we approximate $(1/\rho_{\rm{g}})(\partial P/\partial R)$ as $\sim(c_s^2/H_{\rm{g}})(\Delta \Sigma_{\rm{g}}/\Sigma_{\rm{g}})$. 
Thus, for particles with $St \apgt \alpha \, (\Delta \Sigma_{\rm{g}}/\Sigma_{\rm{g}})^{-1}$, the radial migration towards pressure maximum dominates over turbulent diffusion. In a self-gravitating disc in thermal equilibrium, the amplitude of the spiral structure induced by gravitational instability
has to provide a stress large enough to balance the effective cooling rate (see Eq. \ref{tcool}). Thus,
it is expected that the strength of the modes are proportional to the cooling rate. In this context, by performing high-resolution three-dimensional SPH
simulations of self-gravitating discs evolving under the analytically prescribed cooling
function expressed by Eq. \ref{tcool}, \citet{CossinsLodatoClarke2009} have found that the
strength of the modes varies such that: 
\begin{equation}
\biggl < \frac{\Delta \Sigma_{\rm{g}}}{\Sigma_{\rm{g}}}\biggr >\approx\frac{1}{\sqrt{\beta_{\rm{cool}}}} \, ,
\label{eq:beta_cooling}
\end{equation}
where $\langle \Delta \Sigma_{\rm{g}}/\Sigma_{\rm{g}} \rangle$ indicates the relative RMS amplitude of the surface density perturbations. 
This is closely linked to the fact that smaller cooling times result in larger pressure gradients and thus a more efficient dust trapping into the pressure maxima \citep{Gibbons1}. 
Moreover, the value of the gravoturbulent parameter $\alpha$ depends on the imposed cooling. In detail, using the general results of viscous disc theory (see Section 2 of \citealt{LodatoRice2004}), the value of $\alpha$ needed to balance the imposed cooling varies such that $\alpha\propto \beta_{\rm{cool}}^{-1}$. Thus, the minimum Stokes number above which the advection towards pressure maxima dominates over gravoturbulent diffusion varies with the cooling as $\beta_{\rm{cool}}^{-1/2}$. In other words, the range extension of particle size that show the concentration into pressure maxima decreases with decreasing cooling time due to the combined effect of radial drift and turbulent mixing. %Anyway, as shown in \citet{Gibbons1}, at any given value of cooling time, the maximum density enhancement is mostly due to particles with $St\sim 1$.

In Fig. \ref{img:plot_total} we show the final surface density rendered images of the dust for all whole disc. In order to show the differential distribution from different ranges of dust grain size, the upper panels show the surface density integrated in size in the range $[1\mu m,100\mu m]$ (left) and $[1 cm, 100 cm]$ (right). As expected, the dust density distribution for particles with Stokes number close to unity (see the upper right panel of Fig. \ref{img:plot_total}) show the highest concentration in spiral arms. In this case, the spiral arms are very thin and display a strong contrast between arm and interarm regions, while the structure shown by smaller particles is broader and essentially follows the gas structure (see Fig. \ref{img:plot_density}). %there is a high degree of correlation between the gas surface density structure (shown in Fig. \ref{img:plot_density}) and the arrangements of dust particles. 
%By contrast, as shown in Fig. \ref{img:plot_cut}, the density structure of smaller particle (see the upper left panel of Fig. \ref{img:plot_total}) show much less correlation with the gas surface density structure. As a result, the dust density structure for larger grains is characterized by an higher contrast between arm and interarm regions. 
Note that beyond $\sim 50$ au the density of large particles drops because we have crossed the fragmentation threshold.
The lower panels of Fig. \ref{img:plot_total} show the total dust density distribution (left) and the dust-to-gas ratio (right).
It can be noticed that the dust transport process in a self-gravitating accretion discs quickly modify the radial and azimuthal distribution of the dust-to-gas ratio. Starting from an initial uniform dust-to-gas ratio ($10^{-2}$), dust particles migrate quickly towards pressure maxima leading to a creation of a spiral structure in the dust-to-gas ratio. The local density of particles in interarm regions reaches levels a factor of $\sim10^{-2}$ times that of the initial dust density. It is worth noting that the evolution of the dust-to-gas ratio shows a remarkable depletion of dust grains from the interarm regions since larger particles are more subject to radial drift due to the lower gas density in interarm regions that produce a better coupling between dust and gas (see Eq. \ref{eq:stokes}).

\subsubsection{Influence of fragmentation velocity}

In order to cover a wide range of initial dust density distribution, we run the same simulations, but we now change the value of the fragmentation threshold velocity. As mentioned above, dust aggregates are able to grow by collisions with velocities up to a value that depends on the material properties (\citealt{BlumWurm}, \citealt{Wada}).

In Fig. \ref{img:plot_cut_v1} we show the initial (left) and the final (right) dust density distribution along a slice across the disc using a fragmentation velocity equal to 10 $\rm{m~s}^{-1}$ and an initial size distribution with exponent  $q=3.5$. If the velocity at which particles fragment decreases from $30$ to 10 $\rm{m~s}^{-1}$, grains are not expected to grow to large sizes so that the condition $\rm{St}\sim 1$ is not expected to be met anywhere in the disc. In this case, the efficiency of the dust trapping is reduced since most of the particles are characterised by a Stokes number less than unity. As a result, variations in the dust-to-gas ratio are being smeared out by the turbulent diffusion and the radial drift has only a minor effect in the dust dynamics. 
%As explained above, only particles with $\rm{St}\apgt \alpha \sqrt{\beta}$ (dotted line in Fig. \ref{img:plot_cut_v1}) are subject to radial drift toward pressure maxima.   However, larger particles in interarm regions are more subject to radial drift due to the lower gas density that produce a better coupling between dust and gas (see Eq. \ref{eq:stokes}). Thus, since the Stokes number of larger particles is close to unit in the interarm region, the radial drift produces a migration of larger particles from interarm regions towards pressure maxima. On the other hand, these particles do not exhibit the same level of concentration in wave crests as in previous case due to the lower gas-dust coupling in the pressure maxima.

\subsubsection{Influence of the initial power-law grain size distribution}
\begin{figure}
%\begin{minipage}{\textwidth}
\centering
\includegraphics[scale=0.45]{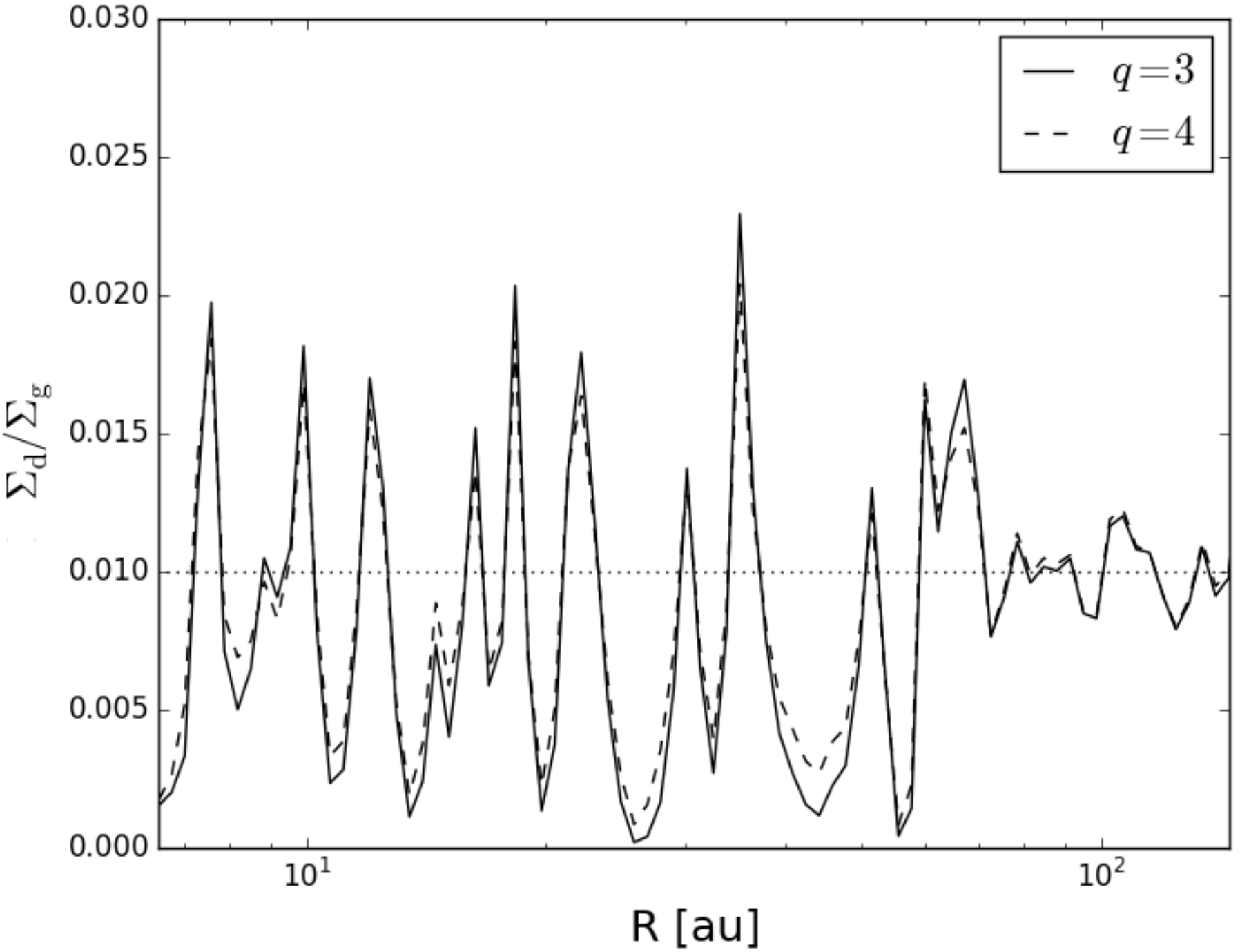}
\caption{Dust-to-gas mass ratio at the end of the simulation ($t=10^3 \,\rm{yrs}$) along a slice across the disc for two values of the power-law exponent of the grain size distribution: $q=3, 4$. The dotted line shows the initial dust-to-gas ratio.}
\label{img:img_d2g}
%\end{minipage}
\end{figure}
As mentioned above, the grain coagulation and fragmentation processes can affect the size distribution quite significantly. 
If the fragmentation process dominates over coagulation, a constant replenishment of smaller grains occurs in the disc producing a steeper size distribution, approaching $n(a)\propto a^{-4}$. In this case, most of the mass will be concentrated in small grains which are less affected by the radial drift toward pressure maxima. By contrast, if the grain coagulation dominates over fragmentation, the resulting grain size distribution is expected to be flatter (e.g. $n(a)\propto a^{-3}$). In this case, the grain coagulation process would produce large grains which migrate quickly toward pressure maxima.
Fig. \ref{img:img_d2g} shows the radial dependence of the final dust-to-gas mass ratio along a slice across the disc for two values of the power-law exponent of the grain size distribution and using a value of the fragmentation velocity equal to $30 \,\rm{m~s}^{-1}$.
For $q=3$, the radial distribution of dust is strongly affected by the migration toward pressure maxima induced by gas drag since most of the dust mass is in large grains. As expected, a steeper size distribution reduce the concentration but only marginally, so that the dust-to-gas ratio in the interarm regions is slightly enhanced with respect to the previous case. 

\subsection{Observational predictions}
\label{sec:obs}
In this section, we present results from simulated ALMA and HiCIAO observations of the sky models computed via RADMC-3D simulations starting from the disc models obtained from hydro simulations.
We translate hydro simulations into model observations using the method explained in Sect. \ref{subsec:montecarlo}. Once having computed the expected full-resolution emission maps, we simulate realistic ALMA and HiCIAO images at (sub-)mm and near-infrared wavelengths.
The synthetic images are analysed in terms of the ability to spatially resolve the spiral features of our models with the aim to test the detectability of gravitationally-induced spiral density waves.

%In order to connect the results from previous simulations to observations we perform radiative transfer simulations and compute simulated ALMA observations at (sub-)mm wavelengths and HiCIAO H-band polarized intensity image. 
%ALMA should ideally provide resolution down to $\sim$ 2 au for discs observed in the Orion star-forming region ($\sim 400\,pc$) and sub-au resolution in the Taurus - Auriga star-forming region ($\sim 140\,pc$). Therefore, the resolution capabilities of ALMA will be sufficient to spatially resolve disc substructure with unprecedented detail with a sensitivity never obtained before \citep{Krumotz}.

\subsubsection{ALMA simulated observations}
\begin{figure*}
\begin{minipage}{\textwidth}
\centering
\includegraphics[scale=0.44]{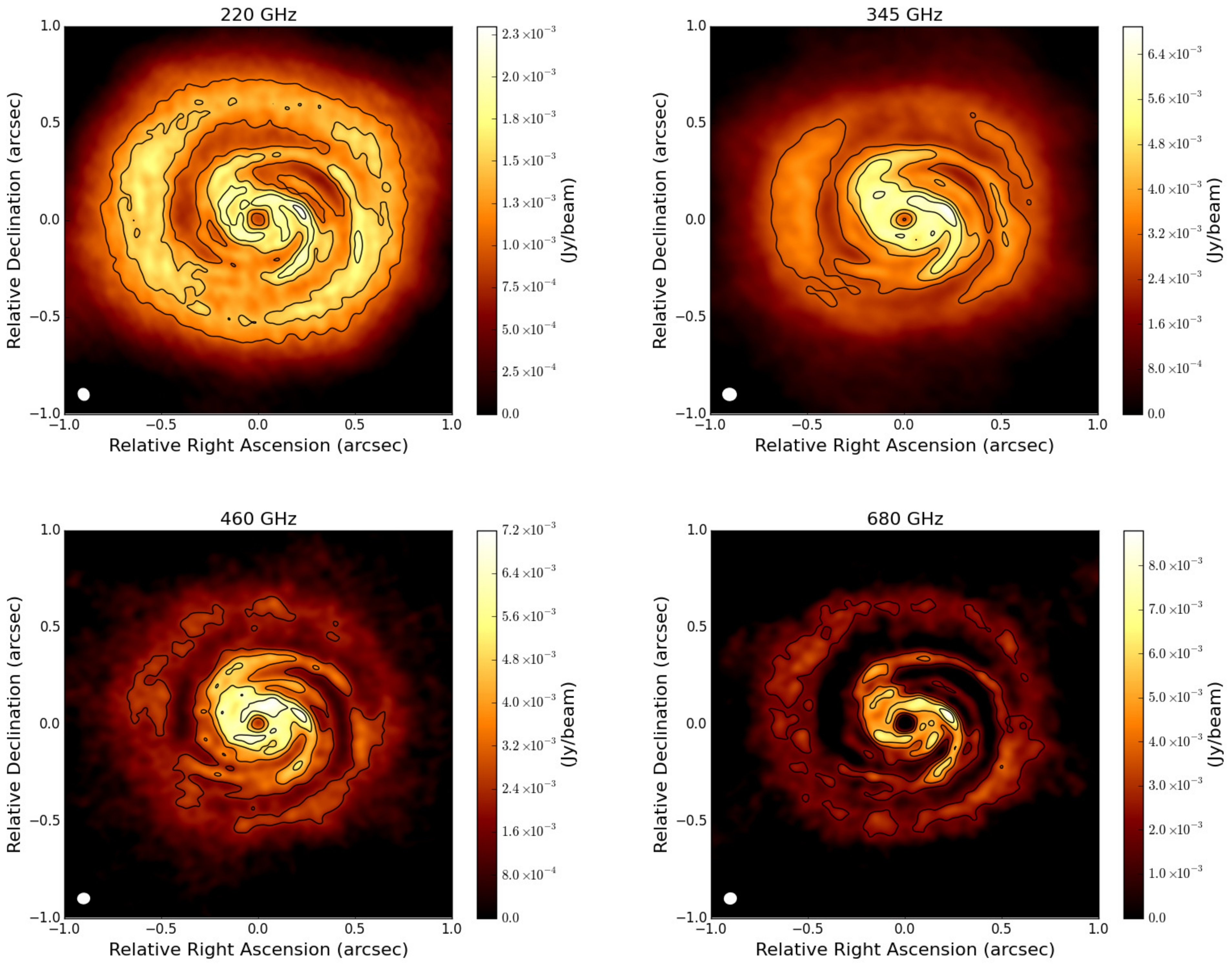}
\caption{ALMA simulated images of the self-gravitating disc model obtained starting from an initial dust grain size distribution with exponent  $q=3.5$ and a fragmentation velocity equal to 30 $\rm{m~s}^{-1}$. Observing frequencies are reported at the top of each image. Contours are 4, 6, 8, 10 $\times$ the corresponding rms noise. The white colour in the filled ellipse in the lower left corner indicates the size of the half-power contour of the synthesized beam (see table \ref{tab:pwv} for details).}
\label{img:alma_0.25_r4_m1_incl_45}
\end{minipage}
\end{figure*}
The resulting emission maps at (sub-)mm wavelength are used as input sky models for the ALMA simulator (version 4.1) in order to produce realistic ALMA observations of self-gravitating disc models. 
The selection of observing parameters is chosen to ensure enough spatial resolution and sensitivity to resolve and detect spiral arms in the intermediate and external region of the discs.  Assuming a perfect calibration of the visibility measurements, the computation of ALMA simulated observations include the effects of thermal noise from the receivers and atmosphere. The signal atmospheric corruption effects into the visibilities measurements is evaluated taking into account the atmospheric transparency computed using the Atmospheric Transmission at Microwaves (ATM) code \citep{Pardo} starting from average site conditions and input values for the ground temperature and the Precipitable Water Vapour (PWV) provided by the user. All the sources are located in the Ophiucus star-forming region ($\sim$ 140~pc) and are observed with a transit duration of one hour. 

First, we focus on a representative disc model obtained from  a given set of model parameters. In Fig.  \ref{img:alma_0.25_r4_m1_incl_45} we show ALMA simulated observations at different observing wavelengths of the disc models obtained starting from an initial dust grain size distribution with exponent  $q=3.5$ and a fragmentation velocity equal to 30 $\rm{m~s}^{-1}$. The observation parameters adopted in the images presented here and the measured total fluxes are summarized in Table \ref{tab:pwv}.
It can be noticed that the spiral structure is readily detectable by ALMA over a wide range of wavelengths. Note that relative intensity between the spiral arms and the external region decreases with increasing frequency. This variation is linked to the optical regime of the spiral arms at different frequencies. It can be shown that, while the interarm regions are optically thin, the arm regions became optically thick for frequencies $\apgt 345$ GHz. Thus, for $\nu \apgt 345$ GHz, the radiation coming from the central protostar is more likely to be captured by the arm regions, producing a fainter emission from the outer regions.
\begin{table}
\setlength{\tabcolsep}{3.6pt}
%\begin{minipage}{\textwidth}
\begin{center}
\begin{tabular}{ccccc}
\midrule
Frequency               & PWV & beam size  & Total flux & rms \\
(GHz)   & (mm) & (arcsec $\times$ arcsec) & (Jy) & (mJy/beam) \\
\midrule
$220$          &    $2.75$ & 0.063 $\times$ 0.055& 0.71 & 0.18\\
$345$          &    $1.26$ & 0.070 $\times$  0.063&0.93&0.49\\
$460$          &    $1.26$ &0.062 $\times$ 0.054&1.12&0.84\\
$680$          &    $0.47$ &0.062 $\times$ 0.057&1.31&1.15\\
\bottomrule
\end{tabular}
\caption{Atmospheric conditions, beam size, total flux and rms for the simulated observations shown in Fig. \ref{img:alma_0.25_r4_m1_incl_45}.}
\label{tab:pwv}
\end{center}
%\end{minipage}
\end{table}

In addition, as already noted by \citet{Dipierro2014}, while the original density structure of the disc is characterized by a relatively large number of arms (see Fig. \ref{img:plot_density}), the ALMA image show a spiral pattern with two spiral arms since the smaller-sized arms have been smeared out by the limited resolution of the observations.
\begin{figure*}
\begin{minipage}{\textwidth}
\centering
\includegraphics[scale=0.44]{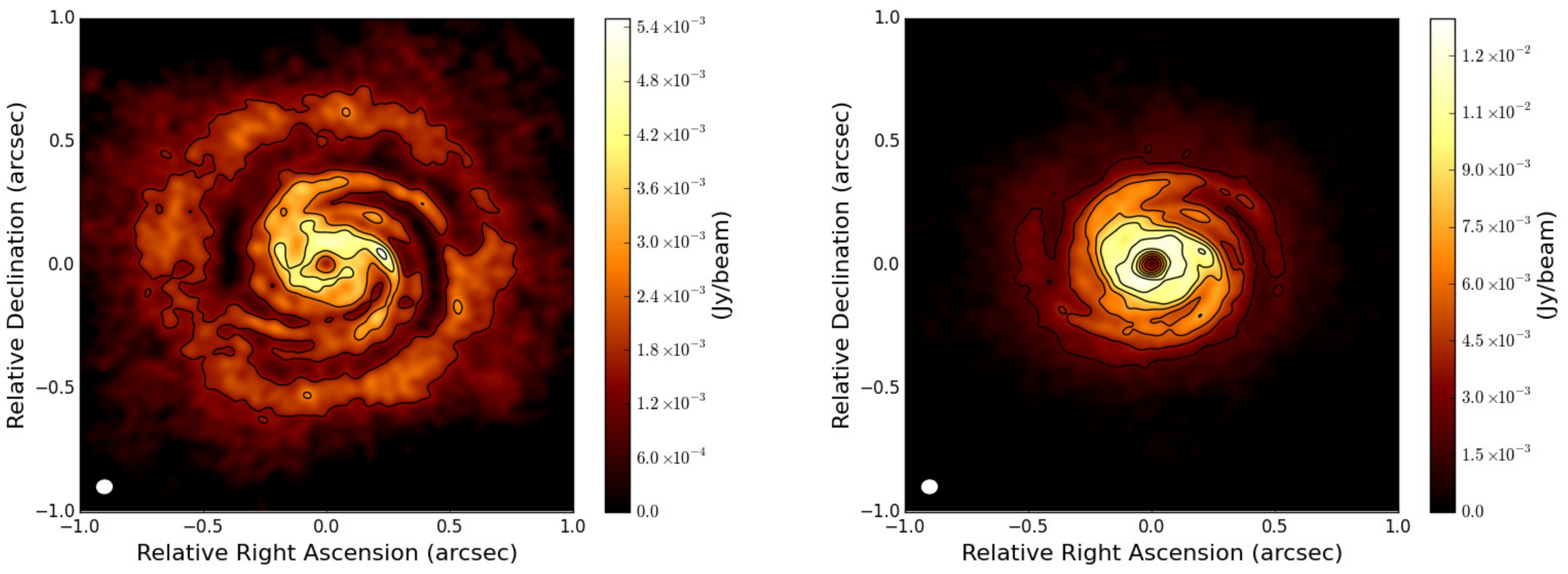}
\caption{Comparison of ALMA simulated images at 460 GHz of disc models obtained from simulations with an initial dust grain size distribution with power-law exponent $q=3$ (left) and $q=4$ (right) and a fragmentation velocity equal to 30 $\rm{m~s}^{-1}$. Contours are 4, 6, 8, 10 $\times$ the corresponding rms noise: 0.78 mJy/beam (left) and 0.72 mJy/beam (right). The white colour in the filled ellipse in the lower left corner indicates the size of the half-power contour of the synthesized beam: 0.062 arcsec $\times$ 0.054 arcsec.}
\label{img:cfr_q}
\end{minipage}
\end{figure*}
\begin{figure*}
\begin{minipage}{\textwidth}
\centering
\includegraphics[scale=0.37]{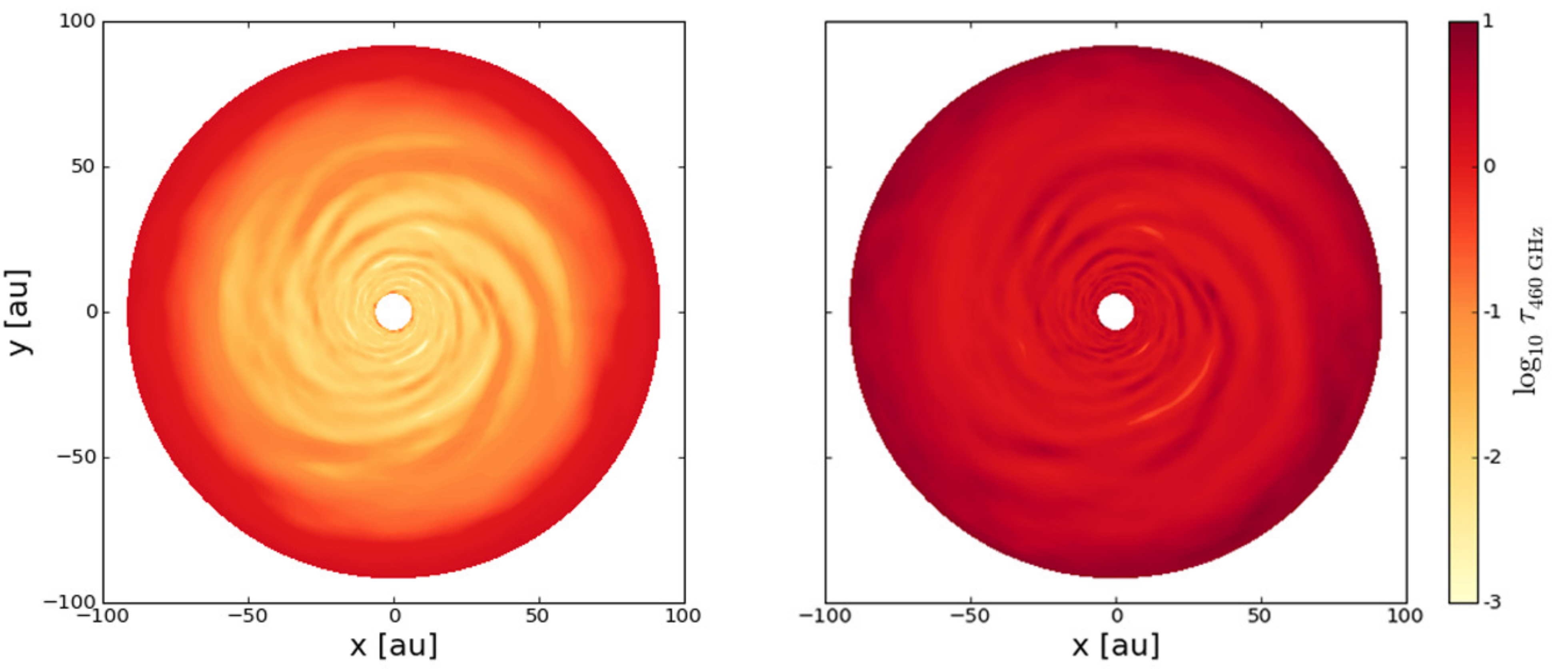}
\caption{Comparison of optical depth maps at 460 GHz of disc models obtained from simulations with an initial dust grain size distribution with power-law exponent $q=3$ (left) and $q=4$ (right).}
\label{img:cfr_tau}
\end{minipage}
\end{figure*}
\begin{figure*}
\begin{minipage}{\textwidth}
\centering
\includegraphics[scale=0.37]{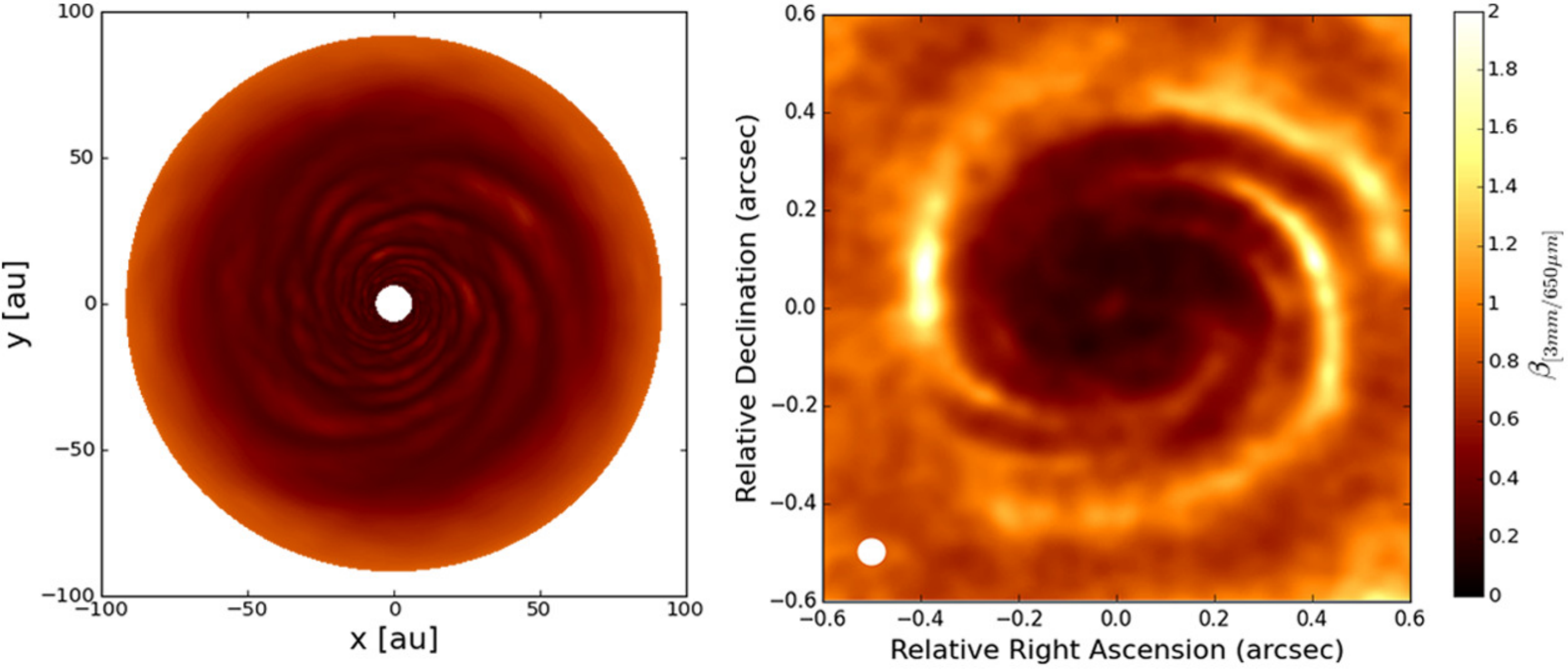}
\caption{The $\beta$ parameter computed from (left) the model data and from (right) two simulated ALMA observations at 650 $\rm{\mu}$m and 3 mm of the disc model obtained starting from an initial dust grain size distribution with exponent  $q=3$ and a fragmentation velocity equal to 30 $\rm{m~s}^{-1}$. The white colour in the filled ellipse in the lower left corner of the right panel indicates the size of the half-power contour of the synthesized beam: 0.054 arcsec $\times$ 0.052 arcsec.}
\label{img:cfr_beta}
\end{minipage}
\end{figure*}

In Fig. \ref{img:cfr_q} is shown the comparison of two ALMA simulated observations at 460 GHz of the disc models obtained from simulations with an initial dust grain size distribution with power-law exponent $q=3$ (left) and $q=4$ (right). As expected, the slope of the dust grain distribution affects the resulting observation. For $q=4$, most of the dust mass is concentrated in small grains. This produces an optically thick inner region that capture all the radiation coming from the central protostar. As a result, due to the remarkable decrease of the contrast between arm and interarm region, the spiral pattern is only barely with ALMA.
To illustrate the effects of varying the slope of the size grain distribution, in Fig. \ref{img:cfr_tau} we show the optical depth maps at 460 GHz of the disc models obtained from simulations with an initial dust grain size distribution with power-law exponent $q=3$ (left) and $q=4$ (right). 
The optical depth is evaluated locally using the following expression:
\begin{equation}
\tau_{\mathrm{\nu}}=\int_{a_{\rm{min}}}^{a_{\rm{max}}} \frac{\partial \Sigma_{\rm{d}}} {\partial a} \kappa_{\mathrm{\nu}}(a) \,d a \, ,
\end{equation}
where $\kappa_{\mathrm{\nu}}(a)$ represents the absorption opacity at frequency $\nu$ for the dust species with size $a$.
It can be noted that, for grain size distribution with $q=3$, the spiral pattern region is optically thin. On the contrary, for steeper grain size distribution the optical regime is thick for all the disc.

Observation at (sub-)millimetre  wavelengths can probe some dust properties in the disc interior. In detail, by making some assumptions on the chemical composition, shape and size distribution of the dust grains, it is possible to put constraints on the level of grain growth. For an optically thin disc in the Rayleigh-Jeans limit, the sub-millimetre spectral energy distribution (SED) can be approximated as  a power-law $F_{\nu} \propto \kappa_{\nu} \,\nu^2\propto\nu^{\alpha}$ where $\alpha$ is the spectral index. For grain size distribution of the form adopted in the present work (see Eq. \ref{grainsizedistrib}), a good approximation for the dust opacity is a power law: $\kappa_{\nu}\propto \nu^{\beta}$, where the parameter $\beta$ is a very good indicator of the level of grain growth. While the typical value for ISM grains is $\beta_{\rm{ism}}=1.7$, several observations (e.g. \citealt{Testi2001,Testi2003}, \citealt{Natta2004}, \citealt{Ricci2010}) have found that $\beta_{\rm{disc}}<\beta_{\rm{ism}}$,  which is naturally interpreted in terms of grain growth (e.g. \citealt{Draine2006}). 
In the context of the present work, we calculate the value of the $\beta$ parameter by computing the ratio of the ALMA simulated images at two different frequencies, with the aim to infer the different level of dust concentration of large grains in arm and interarm regions. The $\beta$ parameter can be obtained as:
\begin{equation}
\beta=\frac{\ln{F_{\nu_1}} - \ln{F_{\nu_2}}}{\ln{\nu_1} -\ln{\nu_2}}-2 \, ,
\end{equation} 
where $F_{\nu_1}$ and $F_{\nu_2}$ are the fluxes at frequencies $\nu_1$ and $\nu_2$, respectively.  In Fig. \ref{img:cfr_beta} we show the map of $\beta$ computed from the model data (left) and the $\beta$ parameter calculated from two simulated ALMA observations at 650 $\rm{\mu}$m and 3 mm (right). We consider the disc model obtained starting from an initial dust grain size distribution with exponent  $q=3$ and a fragmentation velocity equal to 30 $\rm{m~s}^{-1}$. At wavelengths mentioned above we verify that the disc is optically thin and the dust temperature is high enough to ensure that the disc is in the Rayleigh-Jeans limit. Therefore, the  variations of $\beta$ can be interpreted as a sign of dust trapping regions. From Fig. \ref{img:cfr_beta} it can be noticed that the beta index map has a shape that resemble the spiral structure of the disc model. The spiral arms, where most of the centimetre sized grains are concentrated due to the radial drift, are characterized by a lower value of $\beta$. By contrast, the $\beta$ parameter is relatively higher in the interarm regions, where large grains are depleted due to the dust trapping mechanism. Thus, taking into account all the assumptions about the dust grain properties, the dust trapping induced by GI produces a migration of large dust grains into spiral arms region that can be potentially investigated through multi-wavelenght observations.

\subsubsection{Polarized H-band scattered light images}
\begin{figure}
%\begin{minipage}{\textwidth}
\centering
\includegraphics[scale=0.5]{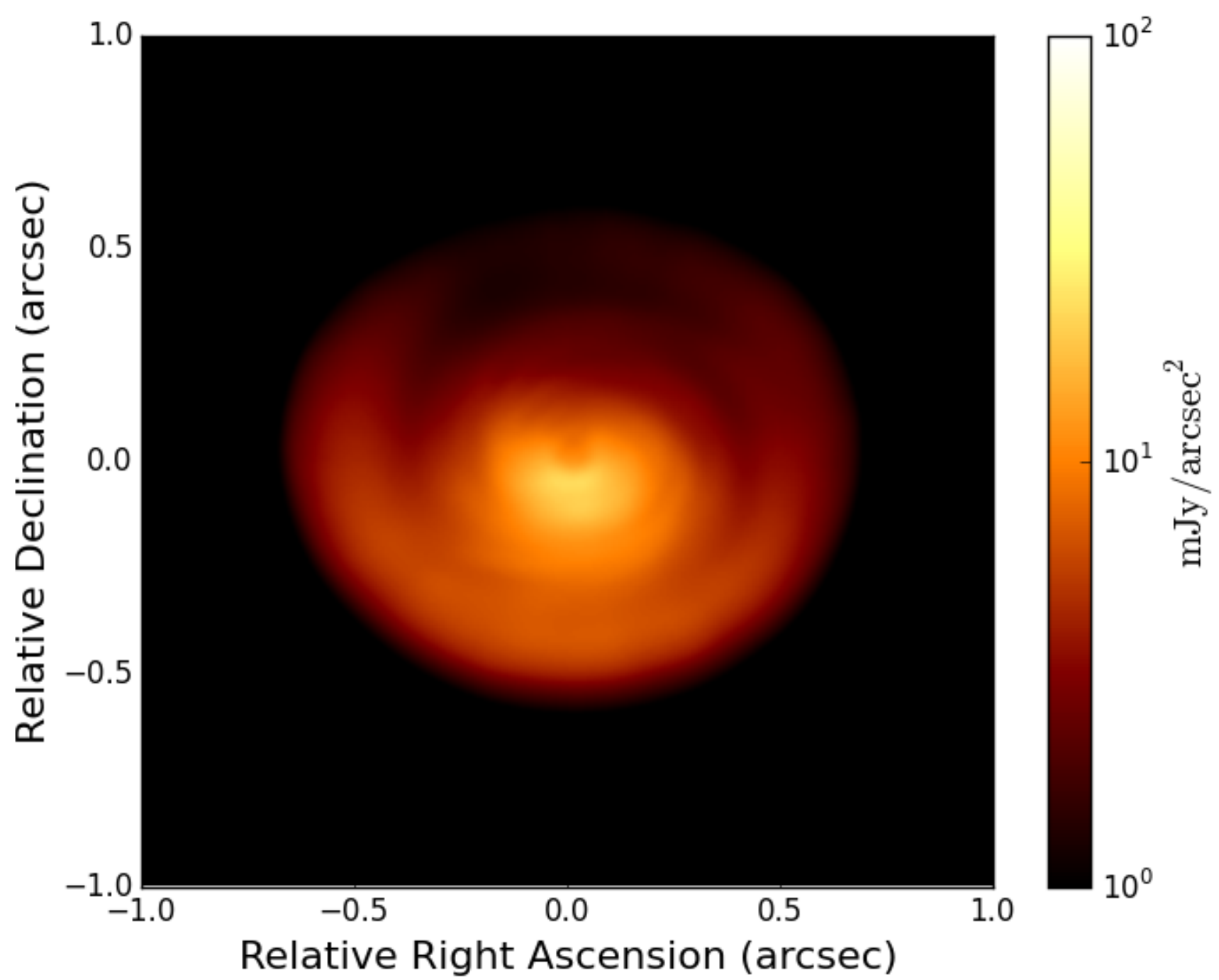}
\caption{Synthetic HiCIAO H-band polarized intensity image of the non-face-on disc model (inclination of 30\textdegree) obtained starting from an initial dust grain size distribution with exponent $q=3.5$ and a fragmentation velocity equal to 30 $\rm{m~s}^{-1}$.}
\label{img:Hband_piccolo}
%\end{minipage}
\end{figure}

In this subsection we show synthetic HiCIAO H-band polarized intensity image of the disc model obtained starting from an initial dust grain size distribution with exponent $q=3.5$ and a fragmentation velocity equal to 30 $\rm{m~s}^{-1}$. As previously mentioned, the polarized intensity images essentially trace the gas disc structure since small dust grains are less subject to the radial drift. The calculated scattered light image convolved with the measured HiCIAO point spread function is shown in Fig. \ref{img:Hband_piccolo}.
The H-band image reveals an asymmetry in brightness between the far (top) and near (low) disc regions. This is due to forward scattering of starlight by large dust particles in a non-face-on disc (inclination of 30\textdegree). In addition, the outer ring-like structure is very bright since the density of smaller particles increases with radius due to the lower value of the fragmentation threshold size in outer disc regions (see Fig. \ref{img:plot_cut}). Moreover, the significative brightness of the outer ring is also caused by the fact that scattered light is mostly determined by the curvature of the disc surface that, due to the flaring of the disc, is very enhanced in the outer regions.
However, due to the higher brightness of the inner regions and the instrument resolution, the spiral structure is not readily detected. The spiral arms and cavities clearly detected in (sub-)millimetre  ALMA images (see Fig. \ref{img:alma_0.25_r4_m1_incl_45}) are smeared out and are thus marginally visible in the HiCIAO H-band image. It is worth remarking that the shape of the HiCIAO PSF is characterized by large sidelobes which produce a decrease of the dynamic range in the convolved images and, consequently, a decrease of the detected contrast between arm and interarm regions. 

\begin{figure*}
\begin{minipage}{\textwidth}
\centering
\includegraphics[scale=0.5]{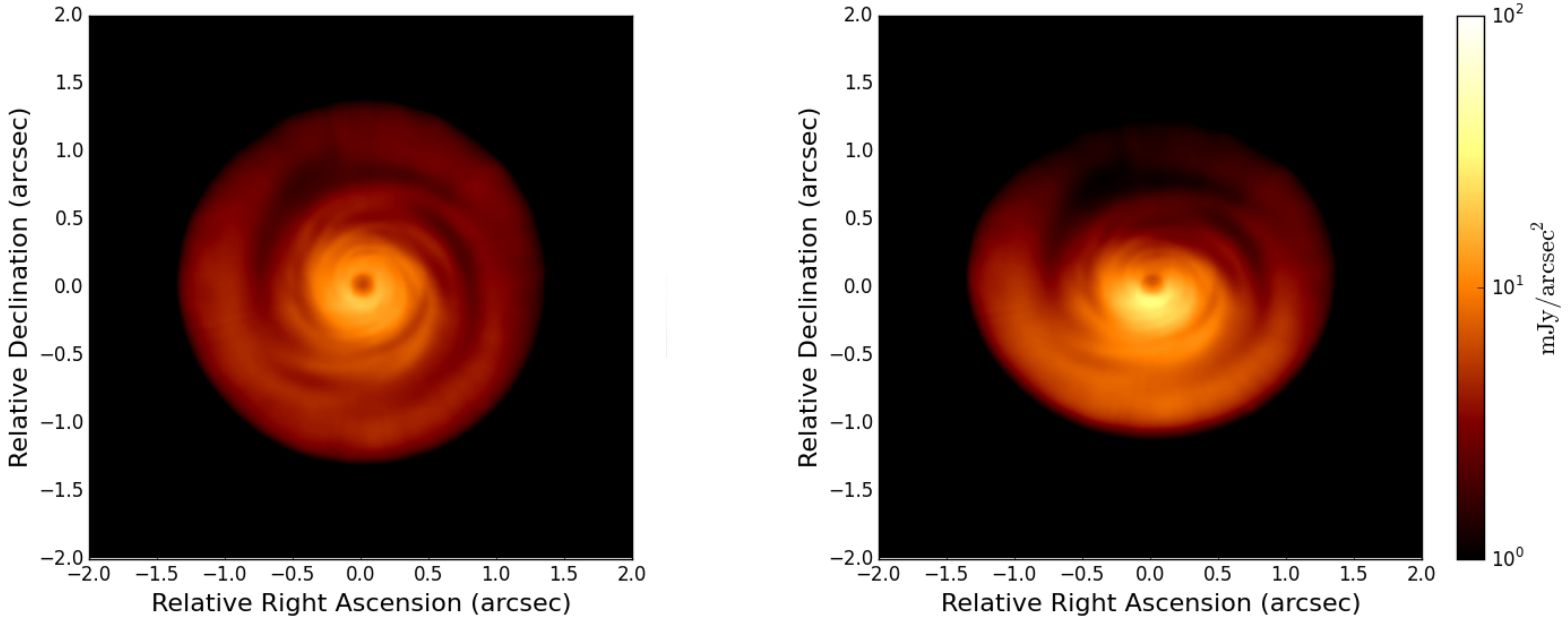}
\caption{Synthetic HiCIAO H-band polarized intensity image of a disc model two times bigger than the model of Fig. \ref{img:Hband_piccolo} (i.e. $R_{\rm{out}}=300$ au) with two different inclinations: 10\textdegree $\,$(left) and 30\textdegree $\,$ (right).}
\label{img:Hband_doppio}
\end{minipage}
\end{figure*}

We also consider a new set of simulations with a gas disc model two times bigger. 
%As expected, we find that the resulting dust density structure of smaller particles at the end of the simulation is not affected by the change of the spatial scale.
In Fig. \ref{img:Hband_doppio} we present the synthetic HiCIAO H-band polarized intensity image of this specific disc model with two different inclinations: 10\textdegree $\,$(left) and 30\textdegree $\,$ (right). A remarkable improvement of the detection of spiral arms can be noticed with comparison to the model previously considered (see Fig. \ref{img:Hband_piccolo}), due to the fact that the HiCIAO PSF is small enough to preserve spiral features in scattering images. As expected, the asymmetry in brightness between the far (top) and near (low) disc regions due to the forward scattering of starlight by large dust particles becomes more prominent with increasing angle of inclination. In addition, since the separation between arm and interarm regions decreases with increasing angle of inclination, the detected spiral structure results sharper for face-on discs. 
Moreover, the contrast between arm and interarm region appears lower in scattering emission than in thermal emission. This is caused mostly by the smaller contrast in dust density of smaller particles with respect to larger particles between arm and interarm regions.
All these effect makes it more challenging to observe the spiral structures in inclined disc through scattering intensity observations.

%we apply a 30 AU diameter circular occulting mask to suppress the bright stellar halo. 

\section{SUMMARY AND CONCLUSIONS}
\label{sec:discuss}
In this paper, by combining hydrodynamical and dust evolution models, we investigate the influence of gravitational instabilities on the dynamics of dust grains. Using representative grains size distributions, we model the dust grains dynamics under the action of systematic and random motion induced by the gas-dust aerodynamical coupling. Once having computed the resulting 2D dust density distribution, we perform 3D Monte Carlo radiative transfer simulation in order to make predictions of self-gravitating discs for future observations. 

The spiral density waves induced by GI produce a remarkable effect on the dust dynamics leading to the formation of significant dust overdensities in the spiral arms regions. The dynamical effect produced by the occurrence of pressure bumps in discs closely depends on the size of the particles. We found that, assuming that particles are able to grow through collisions with velocities up to  10 and 30 $\rm{m~s}^{-1}$, the dust density structure of centimetre sized particles is affected by the presence of the spiral overdensities. While smaller particles closely trace the overall gaseous spiral structure, larger particles experience the highest concentration in density maxima. The influence of the dust trapping mechanism has been tested by computing the dust-to-gas ratio. Since most of the mass is concentrated in the largest grains, the dust transport process modifies the radial and azimuthal distribution of the dust-to-gas ratio leading to a creation of a spiral structure. 

The observational predictions of the resulting models show that the resolution capabilities and sensitivity of ALMA are sufficient to spatially resolve the peculiar spiral structure of gravitationally unstable discs with an acceptable signal-to-noise ratio for non-face-on discs located in the Ophiucus ($\sim$ 140~pc) star-forming region. 
Through multi-wavelenght (sub-)millimetre  observations, we have investigated whether the density enhancement of the larger dust grains into spiral arm regions produces signatures in the spectral index map that can be detected using ALMA. We find that the migration of larger grains into spiral arms region produces clear variations in the spectral index that can be interpreted in terms of different level of grain growth between arm and interarm regions.

In addition, our simulations show that the development of gravitational instability can create strong enough surface density perturbation that could be detected in near-infrared scattered light with HiCIAO.
Therefore, the spiral arms observed to date in protoplanetary disc in polarized scattered light (\citealt{GradyMuto2012}, \citealt{Garufi}, \citealt{GradyMuto2013}) might be the result of spiral density perturbations induced by the development of gravitational instabilities.
%The contrast between arm and interarm regions 
%For such evolved discs, it is more common to explain the detected spiral structures with the dynamical interaction with a third body, like a companion star or an embedded planet. 
There are several scenarios that have been proposed to explain the occurrence of spirals in protoplanetary discs \citep{Boccaletti}. The more widely accepted one is based on tidal interaction of stellar encounter or embedded protoplanets.
%Although gravitational instability is a possible scenario which is most likely to occur in the early stage of star formation, \citet{Fukagawa} and \citet{Christiaens2014} have shown that the outer regions of the transitional disc of the Herbig Fe star HD 142527 is very close to the instability regime. 
However, \citet{Juhasz} have shown that planets by themselves cannot induce density perturbation that can be observed in near-infrared scattered light with current telescope for sources at the distance comparable to Ophiucus star-forming region ($\sim$ 140~pc). Therefore, the development of gravitational instability can be considered a valid theoretical explanation of the spiral structure observed to date in protostellar disc. Moreover, in contrast to the spiral density perturbations due to planet-disc interaction (\citealt{Juhasz}, \citealt{Dong}), the gravitationally-induced density waves has a remarkable effect on the dust dynamics and produce clear observational signatures at (sub-)millimetre  wavelengths.

% it remains unclear how to theoretically explain the spiral structure detected in protoplanetary disc.
%They have estimated, through analytic models, that a surface density perturbation ($\delta \Sigma /\Sigma$) of a factor of $\sim$3.5 or higher above the background disc is required to detect the spiral density waves with HiCIAO for sources at the distance comparable to Ophiucus star-forming region (140 pc). The gas disc model adopted here presents such a 

The simulations presented here contain a number of simplifications that need to be addressed.  
First, since the lifetime of the individual spiral features in self-gravitating discs is of the order of $\Omega^{-1}$, the approximation of a stationary gas density structure for as long as one outer orbital period appears to be simplistic \citep{ClarkeLodato2009}.
Moreover, in this work we neglect the grain growth and fragmentation processing that occur in the disc.
This can be considered an acceptable approximation since the timescale of our simulations is shorter than the typical grain growth timescale (\citealt{Brauer2008}, \citealt{Okuzumi2011}) in the intermediate and outer part of the disc.
In addition, it is expected that the occurrence of long-lived non-axisymmetric structures in the disc affects the azimuthal dynamics of dust grains leading to an additional systematic drift motion of the dust particles towards the pressure maximum (\citealt{Nakagawa1986}, \citealt{Birnstiel2013}). Assuming azimuthal and radial trapping simultaneously would lead to stronger concentrations of dust in pressure traps.
%In addition, the effect that large accumulations of particles have on the gas dynamics due to the back reaction of the particles should be taken into account (\citealt{Gibbons1, Gibbons2}). It is expected that there will be regions where the particle density can reach values comparable to the local gas density. In these cases, the drag that the dust exerts on the gas may become more important leading to a reversed dust motion \citep{Nakagawa1986}. Moreover, since most of the dust mass is concentrated in the largest grains, the spiral arms can potentially be regions where the self-gravity of the solid particles might play a crucial role in accelerating planetesimal growth through direct gravitational collapse \citep{RiceLodato2004,RiceLodato2006}.
%Therefore, the dust trapping capability of spiral arms investigated using the one dimensional approach adopted here can be considered a pessimistic estimation.
 %By simultaneously modeling a wide range of grain sizes, our aim is to follow the detailed evolution of the dust in order to investigate how dust evolution is influenced by perturbations induced by GI. 
 The influence of time-dependent density perturbations induced by GI and the drift caused by azimuthal pressure gradients will be the topic of future works.

\section*{ACKNOWLEDGEMENTS}

The authors are grateful to C. P. Dullemond for making RADMC-3D available. We thank the referee for useful suggestions.
We also thank Attila Juhasz for support in solving some problems with RADMC-3D. 
GD acknowledges Leiden Observatory for hosting and providing CPU time in the Sterrewacht workstations.
GD and GL acknowledge financial support from PRIN MIUR 2010-2011, project ``The Chemical and dynamical evolution of the Milky Way and Local Group Galaxies'', prot. 2010LY5N2T. PP is supported by Koninklijke Nederlandse Akademie van Wetenschappen (KNAW) professor prize to Ewine van Dishoeck. This work
was partly supported by the Italian Ministero dell'Istruzione, Universit\`a e Ricerca through the grant Progetti Premiali 2012 - iALMA (CUP C52I13000140001). 

\bibliography{biblio}
\bibliographystyle{agsm}
\end{document}